\documentclass[
 reprint,
 amsmath,amssymb,
 aps,superscriptaddress
]{revtex4-2}
\usepackage{epstopdf}
\usepackage{amssymb}
\usepackage{amsmath}
\usepackage{cancel}
\usepackage{color}
\usepackage{braket}
\usepackage{graphicx}
\usepackage{epstopdf}
\usepackage{verbatim}
\usepackage{siunitx}
\usepackage{soul}
\usepackage{xr}
\usepackage[toc,page]{appendix}
\usepackage[section]{placeins}
\usepackage[utf8]{inputenc}
\usepackage[colorlinks]{hyperref}

\DeclareUnicodeCharacter{2009}{\,} 

\begin{document}

\title{
Optical Entanglement of Distinguishable Quantum Emitters}

\author{D. S. Levonian}
\thanks{These three authors contributed equally.}
\affiliation{Department of Physics, Harvard University, Cambridge, Massachusetts 02138, USA}
\affiliation{AWS Center for Quantum Computing, Pasadena, California 91125, USA}
\author{R. Riedinger}
\thanks{These three authors contributed equally.}
\affiliation{Department of Physics, Harvard University, Cambridge, Massachusetts 02138, USA}
\affiliation{Institut für Laserphysik und Zentrum für Optische Quantentechnologien, Universit\"{a}t Hamburg, 22761 Hamburg, Germany}
\affiliation{The Hamburg Centre for Ultrafast Imaging, 22761 Hamburg, Germany}
\author{B. Machielse}
\thanks{These three authors contributed equally.}
\affiliation{Department of Physics, Harvard University, Cambridge, Massachusetts 02138, USA}
\affiliation{AWS Center for Quantum Computing, Pasadena, California 91125, USA}
\author{E. N. Knall}
\affiliation{John A. Paulson School of Engineering and Applied Sciences, Harvard University, Cambridge, MA 02138}
\author{M. K. Bhaskar}
\affiliation{Department of Physics, Harvard University, Cambridge, Massachusetts 02138, USA}
\affiliation{AWS Center for Quantum Computing, Pasadena, California 91125, USA}
\author{C. M. Knaut}
\affiliation{Department of Physics, Harvard University, Cambridge, Massachusetts 02138, USA}
\author{R. Bekenstein}
\affiliation{Department of Physics, Harvard University, Cambridge, Massachusetts 02138, USA}
\affiliation{Racah Institute of Physics, The Hebrew University of Jerusalem, Jerusalem 91904, Israel}
\author{H. Park}
\affiliation{Department of Physics, Harvard University, Cambridge, Massachusetts 02138, USA}
\affiliation{Department of Chemistry and Chemical Biology,
Harvard University, Cambridge, Massachusetts 02138, USA}
\author{M. Lon\u car}
\affiliation{John A. Paulson School of Engineering and Applied Sciences, Harvard University, Cambridge, MA 02138}
\author{M. D. Lukin}
\email{lukin@physics.harvard.edu}
\affiliation{Department of Physics, Harvard University, Cambridge, Massachusetts 02138, USA}
\date{July 2021}

\begin{abstract}
Solid-state quantum emitters are promising candidates for the realization of quantum networks, owing to their long-lived spin memories, high-fidelity local operations, and optical connectivity for long-range entanglement. However, due to differences in  local environment, solid-state emitters typically feature a range of distinct transition frequencies, which makes it challenging to create optically mediated entanglement between arbitrary emitter pairs. We propose and demonstrate an efficient method for entangling emitters with optical transitions separated by many linewidths. 
In our approach, electro-optic modulators enable a single photon to herald a parity measurement on a pair of spin qubits.
We experimentally demonstrate the protocol using two silicon-vacancy centers in a diamond nanophotonic cavity, with optical transitions separated by 7.4~GHz. Working with distinguishable emitters allows for individual qubit addressing and readout, enabling parallel control and entanglement of both co-located and spatially separated emitters, a key step towards scaling up quantum information processing systems.
\end{abstract} 

\maketitle

Solid-state quantum emitters have recently emerged as promising candidates for the realization of quantum networks. They combine a number of advantageous properties including electronic spin qubits with long coherence times \cite{tyryshkin_electron_2012, sukachev_silicon-vacancy_2017, abobeih_one-second_2018}, fast gates \cite{nguyen_quantum_2019}, access to nuclear qubit registers \cite{bradley_ten-qubit_2019, saeedi_room-temperature_2013}, deterministic qubit fabrication \cite{schofield_atomically_2003, evans_narrow-linewidth_2016, chen_laser_2017}, and accessible operating temperatures \cite{maurer_room-temperature_2012, saeedi_room-temperature_2013}. The most important  challenge in scalable quantum information processing with defect centers involves generating high-fidelity entanglement between spatially separated defects. 

Entanglement mediated by photons stands out in comparison with other promising approaches 
\cite{dolde_room-temperature_2013,petit_universal_2020,rosenfeld_efficient_2020}  
as a unique mechanism for long distance entanglement even across room-temperature environments \cite{bernien_heralded_2013}. Long distance entanglement can be used for quantum repeaters and the creation of quantum networks \cite{briegel_quantum_1998, kimble_quantum_2008, childress_fault-tolerant_2006}. 
Fast and efficient spin-photon gates in solid-state emitters were recently demonstrated by employing cavity quantum electrodynamics (cQED), with integration of color-centers in nanophotonic resonators enabling reproducible, compact, on-chip architectures \cite{nguyen_quantum_2019}.
These advances enabled the demonstration of Bell state measurements on asynchronously arriving photons \cite{bhaskar_experimental_2020}, a key capability of quantum repeater stations.

Despite the rapid progress in this area \cite{janitz_cavity_2020}, the state-of-the-art photonic entanglement schemes are incompatible with the broad distribution of optical transitions commonly exhibited by solid state emitters due to strain variations. 
Using frequency-erasing time-tagging or electro-optical frequency shifting, entanglement of distinguishable memories separated by at most $\sim100$~MHz has been demonstrated \cite{vittorini_entanglement_2014, riedinger_remote_2018}, which falls short of the typical frequency spread of $\sim 5-150$~GHz for emitters encountered in micro- and nanophotonic structures \cite{dibos_atomic_2018, chu_coherent_2014, evans_narrow-linewidth_2016}. While multi-stage quantum frequency conversion could cover this mismatch, its high noise and low efficiency has so far restricted its application to conversion from emitter wavelengths to telecommunication wavelengths for long distance communication \cite{yu_entanglement_2020, maring_photonic_2017}.
Instead, individual quantum emitters with near-identical optical resonances are post selected \cite{ates_post-selected_2009, sipahigil_indistinguishable_2014, evans_photon-mediated_2018}, or the optical detuning is actively compensated \cite{bernien_heralded_2013, machielse_quantum_2019}. In practice, 
however, such schemes have limited scalability, the former due to its low yield, and the latter due to substantial overhead in device complexity.

\begin{figure}
\begin{center}	

  \includegraphics[width=\columnwidth]{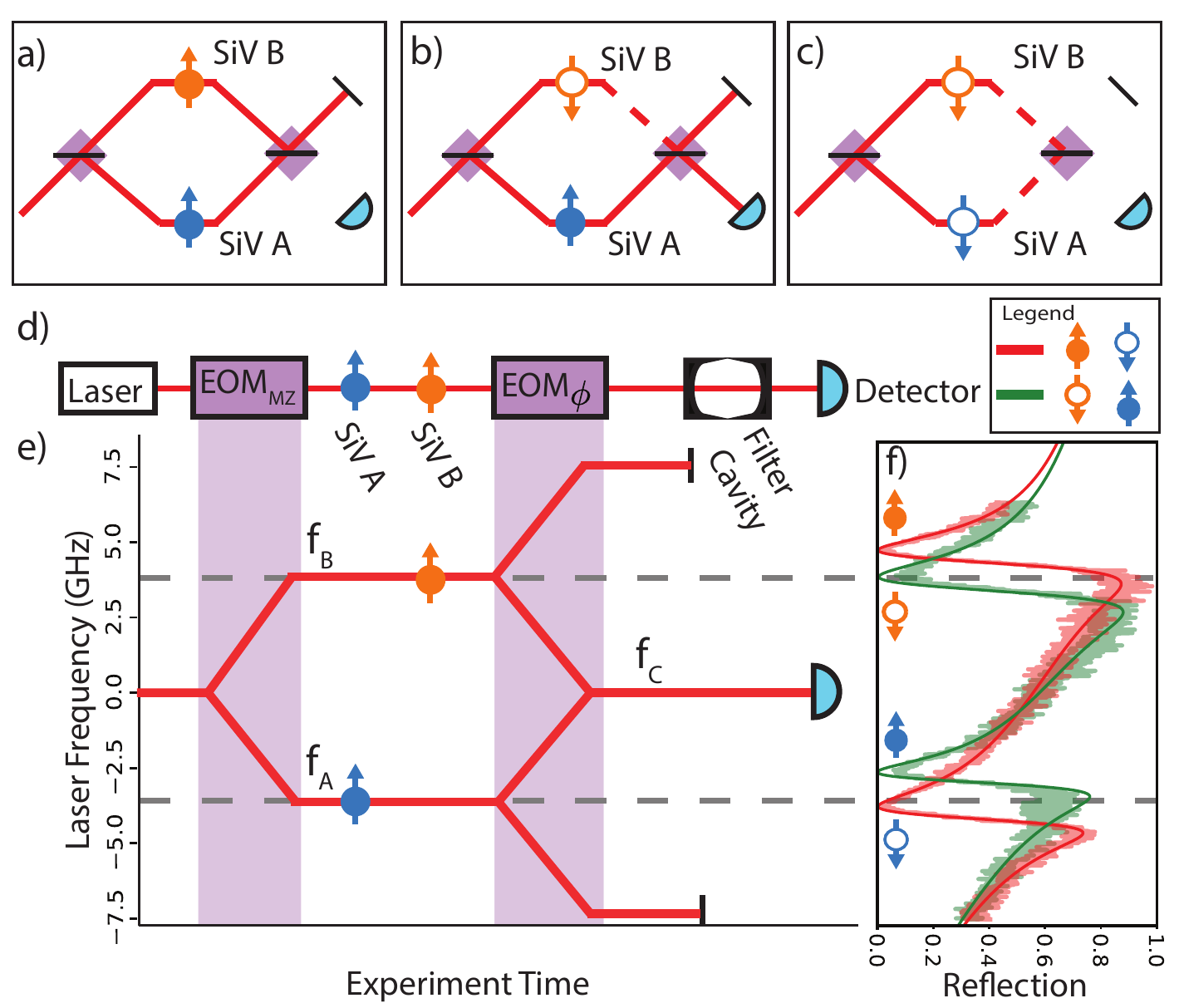}
  \caption{Optical entanglement of distinguishable emitters
  (a) An interferometer is tuned so that when no spin scatters light (full circle) photons leave only through the top port of the interferometer. (b) If just one spin scatters (empty circle) while the other reflects, light is split between the two output ports and the heralding detector receives photons. (c) If both spins scatter, no light leaves the interferometer.
  (d) The physical implementation of the protocol. The two SiVs act as spin-dependent mirrors. The relative phase  between the microwave drive to the two modulators EOM$_\textrm{MZ}$ and EOM$_{\phi}$ sets the phase difference between the interferometer arms.
  (e) The Elitzur-Vaidman Gedanken-experiment implemented as a frequency domain interferometer. The y-axis shows the relative frequency of the relevant photonic modes in the protocol. Two EOMs (purple) play the role of beamsplitters. Detection of photons at the central frequency projects the SiVs into an odd parity state. (f)  The spectrum of the two SiVs under investigation, when initialized in $\left| \uparrow\downarrow\right\rangle$ (green) or $\left| \downarrow\uparrow\right\rangle$ (red).} 
  \label{fig:inhom}
\end{center}
\end{figure}

In this Letter, we propose and demonstrate a scheme to entangle emitters with far-detuned optical transitions which are coupled to an optical cavity. We experimentally realize it using two silicon-vacancy color centers (SiV) in the same diamond photonic crystal resonator, each acting as a spin-dependent scatterer. Our scheme (illustrated in Fig. \ref{fig:inhom}a-c) is inspired by the Elitzur-Vaidman  Gedanken-experiment \cite{elitzur_quantum_1993}. Embedding the two SiVs in the two arms of an interferometer, we use an interaction-free measurement to determine if one (and only one) arm is blocked, determining the joint spin parity by monitoring a dark port of the interferometer \cite{nemoto_photonic_2014, nguyen_integrated_2019}. Unlike the original scheme, our approach uses a frequency domain interferometer (see Fig \ref{fig:inhom}d, e), allowing it to entangle quantum emitters with drastically different optical transition frequencies, and rendering the scheme robust against common noise sources while retaining high resource efficiency. 

The implementation is illustrated in Fig. \ref{fig:inhom}d-e. SiV center A and B are detuned from the nanophotonic cavity such that the system exhibits spin-dependent spectral features of high contrast due to Zeeman splitting of the resonances. The optical transition of SiV A (B) is only resonant with frequency $f_A$ ($f_B$), if the spin is in the $\left|\downarrow\right\rangle_\textrm{\!\tiny{A(B)}}\!$ state, which results in photons being scattered and lost from the interferometer. Otherwise (for $\left|\uparrow\right\rangle_\textrm{\!\tiny{A(B)}}\!$), a Fano interference blocks the light from entering the cavity (Fig. \ref{fig:inhom}f, \cite{nguyen_quantum_2019}), keeping it in the interferometer. 

In each round, the spins of SiVs A and B are first initialized in the state
\begin{equation}
\left|-\right\rangle_\textrm{\!\tiny{A}}\! \otimes \left|+\right\rangle_\textrm{\!\tiny{B}} \propto  \left|\uparrow \uparrow\right\rangle_\textrm{\!\tiny{AB}}\! - \left|\uparrow \downarrow\right\rangle_\textrm{\!\tiny{AB}}\! + \left|\downarrow \uparrow\right\rangle_\textrm{\!\tiny{AB}}\! -\left|\downarrow \downarrow\right\rangle_\textrm{\!\tiny{AB}}\!,
\end{equation}
with $\left|\pm\right\rangle_\textrm{\!\tiny{A(B)}}\! =(\left|\uparrow\right\rangle_\textrm{\!\tiny{A(B)}}\!\pm\left|\downarrow\right\rangle_\textrm{\!\tiny{A(B)}})/\sqrt{2}$, and a photon is prepared in a superposition of two frequency-domain basis states $|f_A\rangle$ and $|f_B\rangle$:
\begin{equation}
\left|\psi\right\rangle_{p,in} = \frac{1}{\sqrt{2}}\left(\left| f_{A}\right\rangle_{p,in}+\left| f_{B}\right\rangle_{p,in}\right).
\end{equation}
This is  achieved by sending a photon at frequency  $f_C=(f_A+f_B)/2$ through an electro-optic intensity modulator (EOM$_{MZ}$) driven at $\omega =(f_B-f_A)/2$ to produce two sidebands at $f_A$ and $f_B$ while suppressing the carrier.
The photon then encounters the two SiVs, where each frequency component is conditionally reflected into the modes described by annihilation operators $\hat{a}$ (for $f_A$) and $\hat{b}$  (for $f_B$).
Next, the two sidebands are recombined, using a phase modulator (EOM$_\Phi$), yielding the mode described by
$\hat{c} = \frac{1}{\sqrt{2}}\left(e^{i\Delta\phi}\hat{a}+\hat{b}\right)$
at frequency $f_C$ ($\Delta\phi$ relative phase). Finally, the light is sent through a filter cavity, which rejects the sidebands, and is detected by a single photon detector (Fig.~\ref{fig:inhom}d).

In case the spins are in the $\left|\uparrow \uparrow\right\rangle_\textrm{\!\tiny{AB}}$ state, both frequency components are reflected, such that the probe photon is in state $\left|\psi\right\rangle_{p} \propto \left| f_{A}\right\rangle_{p}+\left| f_{B}\right\rangle_{p}$ when it arrives at the frequency combiner EOM$_\Phi$, where $\left| f_{A(B)}\right\rangle_p$ indicates a photon in the mode described by $\hat a$ ($\hat b$). We set the interferometer phase $\Delta\phi = \pi$, so that the mode at $f_C$ becomes a dark port of the interferometer, with the amplitudes $\hat{a}$ and $\hat{b}$ interfering destructively. The second EOM transfers the probe photon to the modes at $f_C\pm 2\omega$ (Fig. \ref{fig:inhom}a), where it is rejected by the filter cavity. 
In case of the $\left|\downarrow \downarrow\right\rangle_\textrm{\!\tiny{AB}}\!$ state, there is no photon reflection at either  $f_A$ or $f_B$ (Fig. \ref{fig:inhom}c), also resulting in no events at the detector.
For $\left|\uparrow \downarrow\right\rangle_\textrm{\!\tiny{AB}}\!$ and $\left|\downarrow \uparrow\right\rangle_\textrm{\!\tiny{AB}}$, only one of the frequency components is blocked, destroying the interference condition at the final frequency beamsplitter and allowing the photon to pass through the interferometer (Fig. \ref{fig:inhom}b), revealing the spin parity. 

Similar to the Elitzur-Vaidman Gedanken-experiment, transmission of the photon implies that it did not encounter the scatterer, but nonetheless reveals the scatterer's presence, a phenomenon termed interaction-free measurement.
Importantly, an event at the heralding detector does not reveal which frequency-path was blocked, as the photon could originate from either component of the spin-photon state: 
$\left|\psi_{out}\right\rangle_{\textrm{\!\tiny{AB}},p} \sim -\left|\uparrow \downarrow\right\rangle_\textrm{\!\tiny{AB}}\! \otimes \left| f_A \right\rangle_p + \left|\downarrow\uparrow \right\rangle_\textrm{\!\tiny{AB}}\! \otimes \left| f_B \right\rangle_p.$
A detection event in mode $\hat c$ thus projects the spins to a maximally entangled Bell state: 
\begin{equation} \left|\Psi^+\right\rangle_{\textrm{\!\tiny{AB}}}  =
 \frac{\left|\uparrow \downarrow \right\rangle_{\textrm{\!\tiny{AB}}\!}+ \left|\downarrow \uparrow\right\rangle_{\textrm{\!\tiny{AB}}}}{\sqrt{2}} \end{equation}

This interferometric protocol is both resource efficient and robust. Specifically, a single photon detection is sufficient to herald entanglement, in contrast to two-photon schemes \cite{barrett_efficient_2005}. Nevertheless,
as both frequency components of the photon travel on a common path and a common polarization, it is robust to phase fluctuations of the fiber, requiring no active stabilization of the interferometer. 

\begin{figure}
\begin{center}	
  \includegraphics[width=.8\columnwidth]{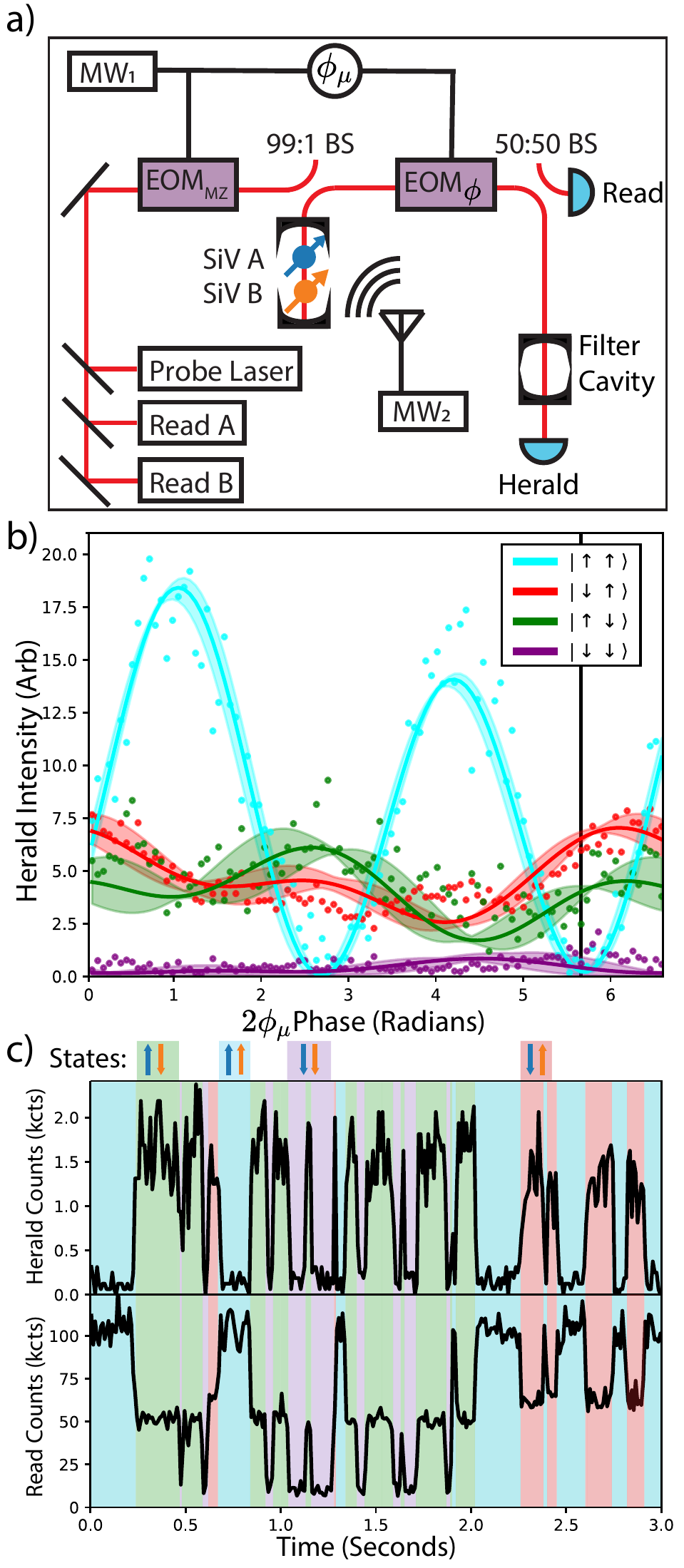}
  \caption{(a) Detailed experimental implementation. The relative phase $\phi_\mu$ between the microwave drive (MW$_1$) to the two modulators EOM$_\textrm{MZ}$ and EOM$_{\phi}$ sets the phase difference between the interferometer arms $\Delta\phi = 2\phi_\mu$. Readout is done sequentially with lasers at $f_A$ and $f_B$ by detecting a fraction of the light before the filter cavity. (b) Transmission to the heralding port vs interferometer phase for all SiV states  $\left|\downarrow\uparrow\right\rangle$ (red circles), $\left| \uparrow\downarrow\right\rangle$ (green), $\left| \uparrow\uparrow\right\rangle$ (cyan) and $\left| \downarrow\downarrow\right\rangle$ (purple). 
  Transmission predicted by a fit of the spin dependent reflection spectrum (Fig \ref{fig:inhom}b,c) as solid lines with variance due to spectral diffusion given by shaded area. Phase and scaling are obtained by fitting the $\left| \uparrow\uparrow\right\rangle$ state. The black vertical line indicates the phase used to collect the entanglement data. (c) Quantum jumps: Transmission through the filter cavity (top panel) and readout port (bottom panel) vs time with the entanglement heralding laser applied continuously. The filter transmission is a spin parity measurement, with high transmission corresponding to odd parity ($\left|\uparrow\downarrow\right\rangle$ or $\left| \downarrow\uparrow\right\rangle$) highlighted with green or red background. Low transmission indicates either $\left|\uparrow\uparrow\right\rangle$ (blue background) or $\left|\downarrow\downarrow\right\rangle$ (purple background).}
  \label{fig:2}
\end{center}
\end{figure}

\noindent Conventional commercial EOMs and signal generators suffice for generating entanglement between emitters separated by up to $|f_A-f_B|\leq80$ GHz in the visible and near infrared wavelength range. This range can be extended to 160~GHz by selecting higher order EOM sidebands with spectral filters. This covers the majority of the inhomogeneous distribution of various color centers in nanostructures, such as C:SiV$^-$, \cite{evans_narrow-linewidth_2016}, YSO:Er$^{+3}$ \cite{dibos_atomic_2018}, and YSO:Nd$^{+3}$ \cite{zhong_optically_2018}.

Our experimental implementation (Fig. \ref{fig:2}a) utilizes a pair of SiV$^-$ centers (A and B) with optical transitions separated by 7.4 GHz, located in the same nanophotonic cavity \cite{nguyen_quantum_2019} with cooperativities $C_A=14.4(1)$ and  $C_B=6.1(1)$, respectively \cite{janitz_cavity_2020}. The cavity is coupled to a waveguide, which adiabatically transfers photons into a tapered fiber with an efficiency of $\eta_{wg} = 0.85 \pm 0.03$. The cavity is detuned from the SiV transitions to yield high reflection contrast for both SiV A and B (Fig. \ref{fig:inhom}f, see SI for details). 
A magnetic field of $B\sim0.45$~T is applied along the common symmetry axis of both SiVs to split the spin conserving optical transitions (with probability of spin changing transition  $r\sim2.3\cdot 10^{-4}$ per cycle). 

To read out SiV A (B), we inject photons at frequency $f_A$ ($f_B$)
and detect them with a superconducting nanowire single photon detector placed before the filter cavity (see \ref{fig:2}a). 
This allows for independent readout of both spin states with fidelity $\mathcal{F}_{\textrm{R},A}=0.9984(1)$ and $\mathcal{F}_{\textrm{R},B}=0.9991(1)$ (see SI). Moreover, the gyromagnetic ratio of SiVs depends significantly on strain, allowing for individual microwave addressing of emitters with the same orientation. Here, we find Zeeman splitting of the ground state spin states of $\omega_{ZA}=12.285$~GHz and $\omega_{ZB}=12.627$~GHz 
(see SI), allowing feedback-based initialization of the individual spins. 

The spins are sequentially initialized, via detection of their state and application of a local rotation to each qubit with a resonant microwave pulse to prepare the state $\left|-+\right\rangle_\textrm{\!\tiny AB}$. Without optical input, we find that an interleaved Hahn-Echo sequence on both spins with pulses separated by $\tau_1 = 412~$ns and $\tau_2 = 423~$ns respectively recovers the initial two-spin-state with a fidelity of $\mathcal{F}_{\textrm{HE},AB}=0.93$, consistent with the corresponding individual Hahn-Echo fidelities $\mathcal{F}_{\textrm{HE},A}=0.96$ and $\mathcal{F}_{\textrm{HE},B}=0.97$\footnote{This indicates that the cross-talk between the microwave pulses is not causing substantial decoherence, as expected from the detuning $\omega_{ZA}-\omega_{ZB}$.}. 
We note that due to drifts in qubit frequencies, the fidelity is reduced during long measurements (resulting e.g. in average $\left\langle\mathcal{F}_{\textrm{HE},AB}\right\rangle=0.85$ over 3 days of measurements). 

We tune the phase of the frequency-bin interferometer by initializing the spins in $\left|\uparrow\uparrow\right\rangle_\textrm{\!\tiny{AB}\!}$ and minimizing the transmission through the interferometer (Fig \ref{fig:2}b, black line).
At the optimal phase, we find the relative transmission rates for the four spin states $T_{\uparrow\uparrow}:T_{\uparrow\downarrow}:T_{\downarrow\uparrow}:T_{\downarrow\downarrow} = 1:14:22:1.2$ (Fig \ref{fig:2}c). The mismatch in reflection between the two odd parity states $\left|\uparrow\downarrow\right\rangle_\textrm{\!\tiny{AB}}$ ($\left|\downarrow\uparrow\right\rangle_\textrm{\!\tiny{AB}}$) is due to interference of the light reflected by the $\left|\uparrow\right\rangle$ state by SiV A (B) with the residual reflection of the $\left|\downarrow\right\rangle$ state of SiV B (A) and the leaked carrier at $f_C$, and can therefore vary depending on their relative phases.

\pagebreak
\onecolumngrid

\begin{figure}[h]
\begin{center}	
  \includegraphics[width=\columnwidth]{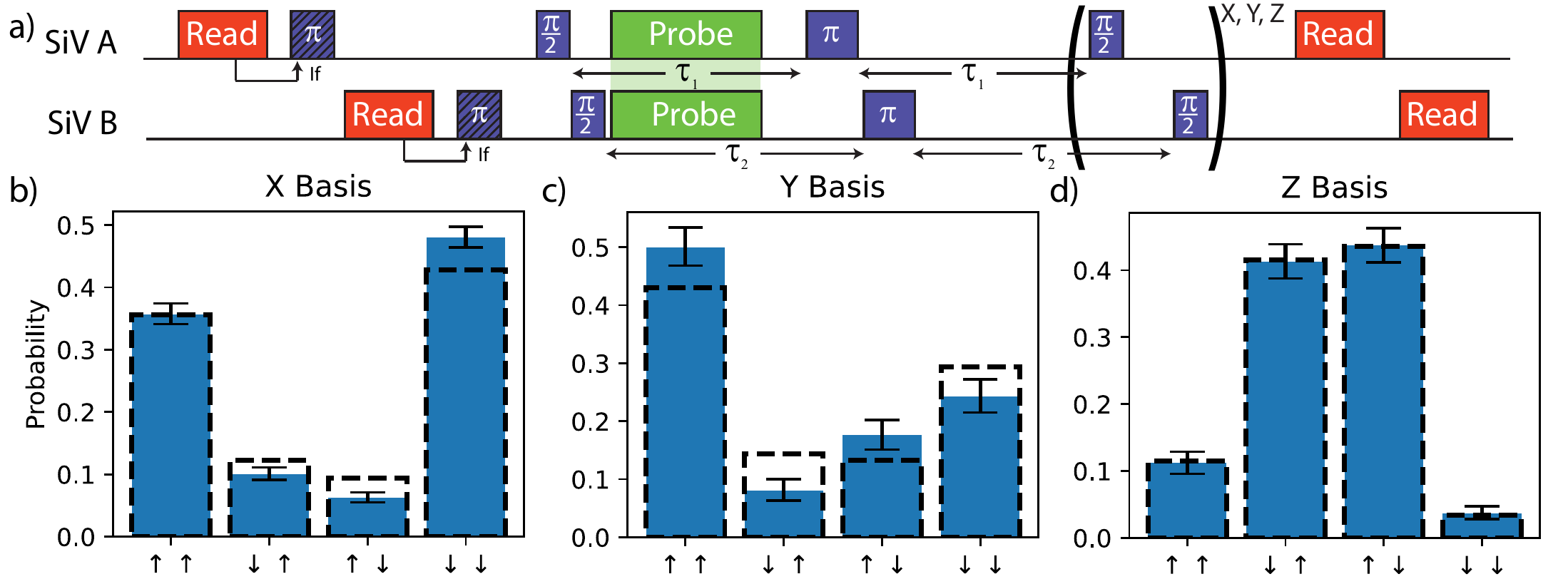}
  \caption{(a) The schematic of the sequence used to entangle the spins. Initialization and readout both apply 10~$\mu$s laser pulses at $f_A$ or $f_B$. Counts are subsequently compared to a threshold to determine the state. Initialization follows this with a conditional $\pi$ pulse. (b,c,d) Correlation statistics of the entangled state in XX, YY and ZZ spin bases. In the experimental data (blue), the measurement was taken with a heralding window of $200$~ns. Dashed black lines are correlations predicted by a theoretical model. Error bars represent $68\%$ confidence interval.}
  \label{fig:3}
\end{center}
\end{figure}
\twocolumngrid
\noindent Similarly, $T_{\downarrow\downarrow}$ is limited by interference of the  finite reflection in the $\left|\downarrow\right\rangle$ states with the leaked carrier. For $T_{\uparrow\uparrow}$ the largest contribution to the finite reflection is the spectral diffusion of the two SiV features and the resulting fluctuation in the phase of the reflected light. 

We entangle the spins by sending a weak coherent pulse with an expected photon number of $0.1$ at the cavity into the interferometer, striking a balance between success probability and decoherence induced by the scattering of extra, undetected heralding photons. 
When a photon is detected in transmission of the filter cavity, this heralds that the spins were prepared in an entangled state. 

To characterize this state, we sequentially measure the 
correlations of the spins of SiV A and B in the X-, Y-, and Z-basis (see Fig \ref{fig:3}b-d). This results in a measured fidelity of 
\begin{equation}
\mathcal{F}_{\left|\Psi^+\right\rangle} = (2p_{\uparrow\downarrow}+2p_{\downarrow\uparrow}+K_{XX}+K_{YY})/4= 0.71(2)
\end{equation}
where $K_{BB}=p_{++}+p_{--}-p_{+-}-p_{-+}$ is the contrast for basis $B=X,Y$ and $p_{ab}$ is the probability for measuring the spin of SiV A (B) in $a(b)\in\{+,-\}$ in $X$- and $Y$-basis, respectively $a(b)\in\{\uparrow, \downarrow\}$ in the $Z$-basis. This confirms that the spins are entangled ($\mathcal{F}_{\left|\Psi^+\right\rangle}>0.5$). As alternative measure of entanglement, we obtain a concurrence of $\mathcal{C}\geq0.37(4)$ (see SI).

To understand the limitations of our protocol and the role of imperfections, 
we compare our experimental results to a model based on the spectrum of the cQED system (Fig. \ref{fig:inhom}c, d). Using the complex reflection coefficient at frequencies $f_A$, $f_B$ and $f_C$, we obtain a predicted transmission through the interferometer for all four spins states (Fig. \ref{fig:2}c)\footnote{The phase and scale is
determined by comparing the prediction to the data of the $\left|\uparrow\uparrow\right\rangle_\textrm{\!\tiny{AB}\!}$ state}. 

\setlength{\belowcaptionskip}{-12pt}
\begin{table}

\begin{ruledtabular}
\begin{tabular}{ll}
Entanglement Error Source & Expected Marginal Error\\
\hline
Local Errors \\
\hspace{5mm} Decoherence $T_2^{\ddagger}$ & $10.7^{+0.5}_{-0.8}$~\%\\
\hspace{5mm} Microwave Pulse Errors & $1.5^{+1.6}_{-1.3}$~\%\\
\hspace{5mm} 2-photon events & $5.3\pm 0.2$~\%\\
Heralded state error\\
\hspace{5mm}Systematic detuning$^\dagger$ & $7.5\pm 0.6$~\%\\
\hspace{5mm}Interferometer phase$^\dagger$ & $7.4\pm 0.9$~\%\\
\hspace{5mm}Carrier leakage & $1.6\pm0.6$~\%\\
\hspace{5mm}Spectral diffusion & $0.3\pm0.1$~\%\\
\hspace{5mm}SiV contrast$^\ast$ & $0.7$~\%\\ 
\hline
Total Expected & $33.0^{+1.4}_{-1.7}$~\%\\
\hline
Total Observed& $29.0^{+1.8}_{-1.9}$~\%\\
\end{tabular}
\end{ruledtabular}
\caption{\label{tab:fidelity} Contributions to the entangled state infidelity. Marginal errors correspond to difference in simulated fidelity between the full model and one with individual sources of error eliminated. Systematic uncertainties are dominated by unknown dispersion of microwave pulses. Notes:
$^\ddagger$ Comparison with Fig. \ref{fig:3}c indicates that decoherence is probably overestimated. 
$^\dagger$Errors due to systematic detuning and optimal interferometer phase are highly correlated. $^\ast$Contribution of SiV contrast relates to residual infidelity when all other sources of error are removed from the model.
}
\end{table}

Including local qubit errors and accounting for a phase drift of the carrier, our model predicts the correlations of the heralded state (see Fig. \ref{fig:3}b-d), and a fidelity of $\sim 0.67\pm0.014$ (see Tab. \ref{tab:fidelity}). The systematic uncertainty stems mostly from microwave dispersion. The largest contribution to the infidelity is spin decoherence, likely caused by the high density of defects in the crystal. Comparison with the experimental data (Fig. \ref{fig:3}b-d) indicates that the model slightly overestimates the impact of spin decoherence (see SI for details).

By eliminating the state preparation and measurement errors, our model estimates the fidelity of the entanglement operation itself to be $\mathcal{F}_\textrm{corr}\sim 0.83$.
Assuming fine tuning of optical parameters, 
no microwave cross-talk, and the best previously observed spin coherence \cite{bhaskar_experimental_2020}, an entanglement fidelity of $\mathcal{F}_{\left|\Psi^+\right\rangle} \sim 0.95$  should be achievable, still limited by residual spin decoherence.
The entanglement rate is currently limited by low detection efficiency ($\eta = 0.04$) and the use of a weak coherent state as heralding state. Together, this yielded a success probability of $6\cdot10^{-4}$ per attempt and an entanglement rate of $0.9$~Hz.
Ultimately this protocol can reach $25\%$ entanglement probability using single photon sources and critically coupled cavities. 
Using spin dependent phase flips in overcoupled cavities \cite{tiecke_nanophotonic_2014}
close to $50\%$ entanglement probability can be reached, resulting in an entanglement rate of 50 kHz and  providing an efficient 
mechanism for quantum networking. 

In summary, we have described a protocol to entangle quantum memories with far-detuned optical transitions, and demonstrated it using to SiVs separated by 7.4~GHz. The protocol is inherently efficient and stable, as it relies on single photon interference in a common-path. 
Our approach can be extended both to spatially separated qubits as well as other spectrally inhomogeneous qubits. The current limits can be circumvented by using stable SiV centers in separate devices, and high entanglement fidelities are possible with previously demonstrated parameters \cite{bhaskar_experimental_2020}. %
We further note that this protocol can potentially result in very high entanglement rates with low loss modulators, integrated filters, and a single photon source instead of weak coherent pulses, opening the door for a broad range of new applications in quantum networking and quantum information processing.  

We thank Pavel Stroganov, Eric Bersin, Leigh Martin  and Neil Sinclair for discussions, Vikas Anant from PhotonSpot for providing SNSPDs, and Jim MacArthur for assistance with electronics. This work was supported by the NSF, CUA, DoD/ARO DURIP, AFOSR MURI, ONR MURI, ARL, and a Vannevar Bush Faculty Fellowship. Devices were fabricated at Harvard CNS, NSF award no. 1541959. M. K. B. and D. S. L. acknowledge support from an NDSEG Fellowship. R. R. acknowledges support from the Alexander von Humboldt Foundation and the Cluster of Excellence 'Advanced Imaging of Matter' of the Deutsche Forschungsgemeinschaft (DFG) - EXC 2056 - project ID 390715994. B. M. and E. N. K. acknowledge support from an NSF GRFP.

\end{document}


\widetext

\begin{center}
\textbf{\large Supplemental Materials: Optical Entanglement of Distinguishable Quantum Emitters}
\end{center}
\setcounter{equation}{0}
\setcounter{figure}{0}
\setcounter{table}{0}
\setcounter{page}{1}
\makeatletter
\renewcommand{\theequation}{S\arabic{equation}}
\renewcommand{\thefigure}{S\arabic{figure}}
\renewcommand{\thetable}{S\arabic{table}}


\section{Experimental Implementation}

\begin{figure*}
\begin{center}	

  \includegraphics[width=\textwidth]{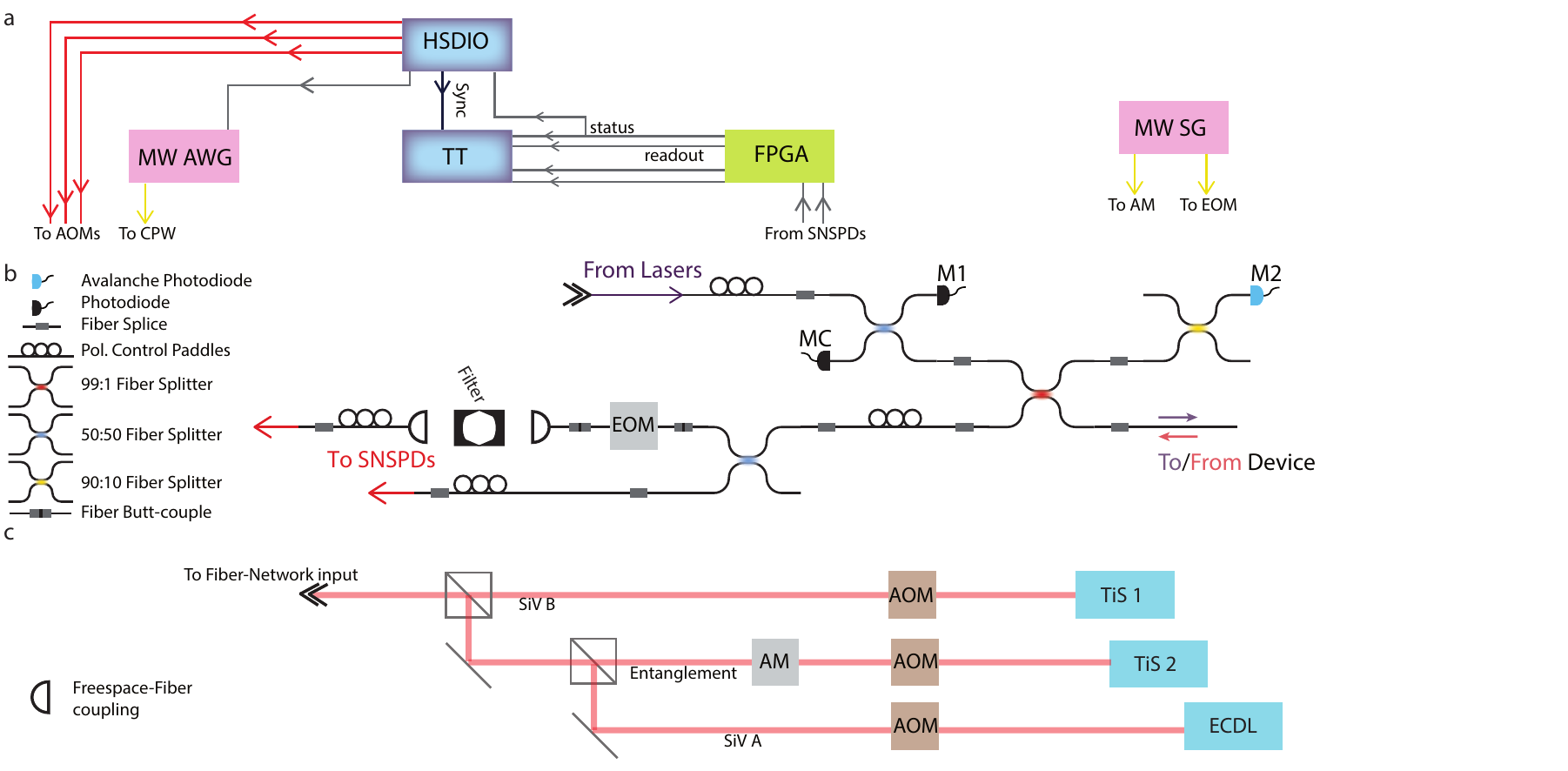}
  \caption{ a Control flow of experiment. All signals are recorded on a time-tagger (TT, PicoQuant HydraHarp 400). b, Fiber network used to deliver photons to and collect photons from the device, including elements for polarization control and diagnostic measurements of coupling efficiencies. c, Preparation of optical fields.}
  \label{fig:setup}
\end{center}
\end{figure*}

\begin{figure*}
\begin{center}	
  \includegraphics[width=\textwidth]{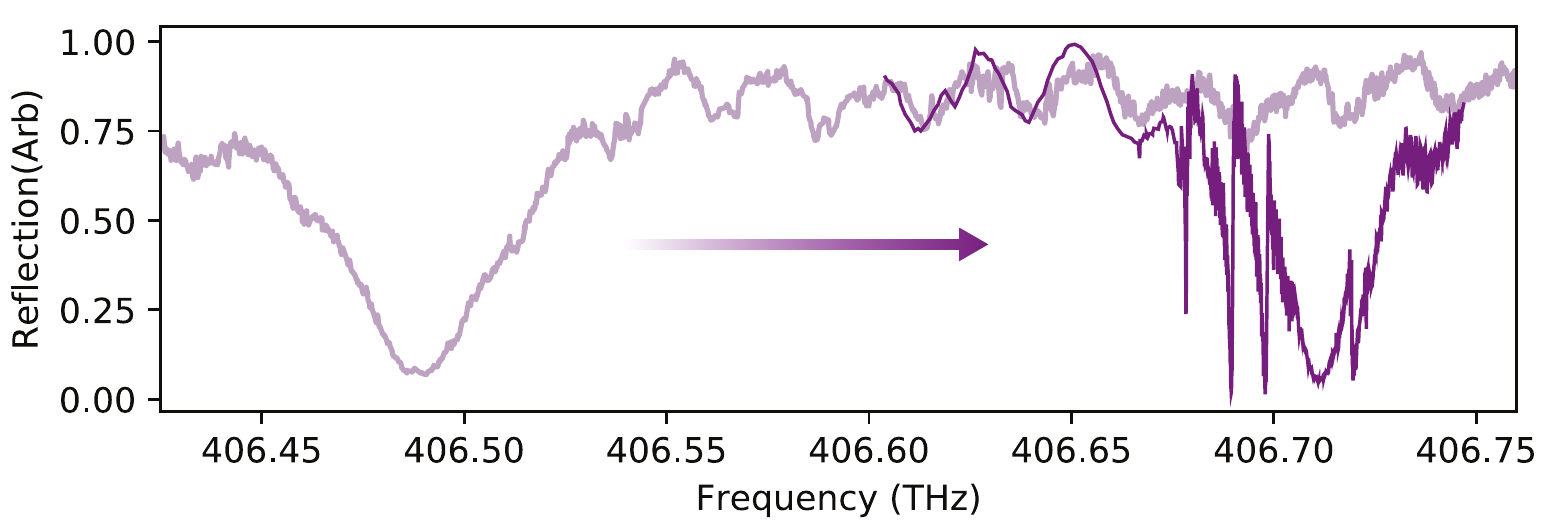}
  \caption{Spectrum before and after cavity tuning.}
  \label{fig:backtuning}
\end{center}
\end{figure*}

Experimental setup and device fabrication 
for millikelvin nanophotonic cavity QED experiments with SiV centers are thoroughly described in separate publications \cite{Burek2014, Burek2017, Atikian2017, nguyen_quantum_2019}. The specific setup is briefly summarized here (Fig \ref{fig:setup}): we perform all measurements in a dilution refrigerator (DR, BlueFors BF-LD250) with a base temperature of 20 mK. The DR is equipped with a superconducting vector magnet (American Magnets Inc. 6-1-1 T), a home-built free-space wide-field microscope with a compact asphere objective (Newport 5722-B-H), piezo positioners (Attocube ANPx101 and ANPx311 series), and fiber and microwave (MW) feedthroughs. Tuning of the nanocavity resonance is performed using a $N_2$ gas condensation technique \cite{Evans2018} (Fig \ref{fig:backtuning}). The SiV-cavity system is optically interrogated through the fiber network without any free-space optics \cite{Bhaskar2019}. The operating temperature of the memory node during the entanglement measurements was $T\sim100-150$~mK.

The experimental apparatus for entanglement has three parts: (1) a system for reading the state of the qubits, (2) a microwave setup for single qubit rotations, and (3) the entanglement heralding system. The readout of the qubits is done with two lasers, an external cavity diode laser (ECDL, Newport Velocity TLB-6711) and a Ti:Sapphire laser (TiSaph, M Squared
SolsTiS-2000-PSX-XF). These two lasers are tuned to the highest contrast frequencies for the two SiVs and sent to the nanophotonic cavity via fiber. Reflected photons are collected via the same fiber and proceed via a directional coupler to a 50/50 beam splitter, one port of which is connected to superconducting nanowire single photon detectors (SNSPD, Photon Spot). Counts collected by this SNSPD are used to determine the state of the SiVs. All detected photons are processed digitally on a field programmable gate array (FPGA, Fig. S1a), and the arrival times of these photon are recorded on a time-tagger (Hydarharp Time Tagger (TT), Fig. S1a). At the end of the experiment, a 10 µs pulse from the readout path is reflected off the device, and photons are counted in order to determine the spin state depending on the threshold.

Rotations on the SiVs are performed by driving magnetic dipole transitions of the SiV electron spins. Microwave signals are produced by a  arbitrary waveform generator (Tektronix AWG70001a 50  GS/s) passed through a  11.3-13 GHz bandpass filter (Marki Microwave FB-1215) to remove digital switching noise and amplified by a  microwave amplifier (MiniCircuits ZVA-183-S+). A DC block prevents any DC current flow into the experiment. The microwaves pass into the dilution refrigerator via stainless steel coaxial cables and are connected to a PCB that is wirebonded to a coplanar microwave waveguide on the surface of the diamond.

The heralded entanglement setup consists of a laser, two electro optic modulators (EOMs) that generate and then combine the sidebands that interact with the SiVs and a tunable Fabry-Perot frequency filter. Light is generated by a second Ti:Sapphire laser (M Squared SolsTiS-2000-PSX-XF) and passes through an  amplitude modulator EOM (EOspace AZ-OK5-10-PFA-PFA-637) before being combined with light from the readout lasers. After reflecting off of the nanophotonic cavity, the herald light is collected at the other port of the 50/50 beam splitter (Evanescent Optics) and is launched onto a free space optics setup by a collimator (Thorlabs PAF-X-2B). The light passes through at telescope consisting of a achromat (LA1509B, focal length f=100mm ) and a f=30mm achromat (AC127-030-B-ML) which mode matches it to the filter cavity. The filter cavity consists of two $99$\% reflective mirrors with a radius of curvature of 5 meters (LayerTec), separated by a 2mm thick  ring piezo (Thorlabs PA44LEW) for tuning. Light is coupled back into a fiber and guided to a second SNSPD by a symmetric set of lenses and collimators.

\begin{figure*}
\begin{center}	

  \includegraphics[width=\textwidth]{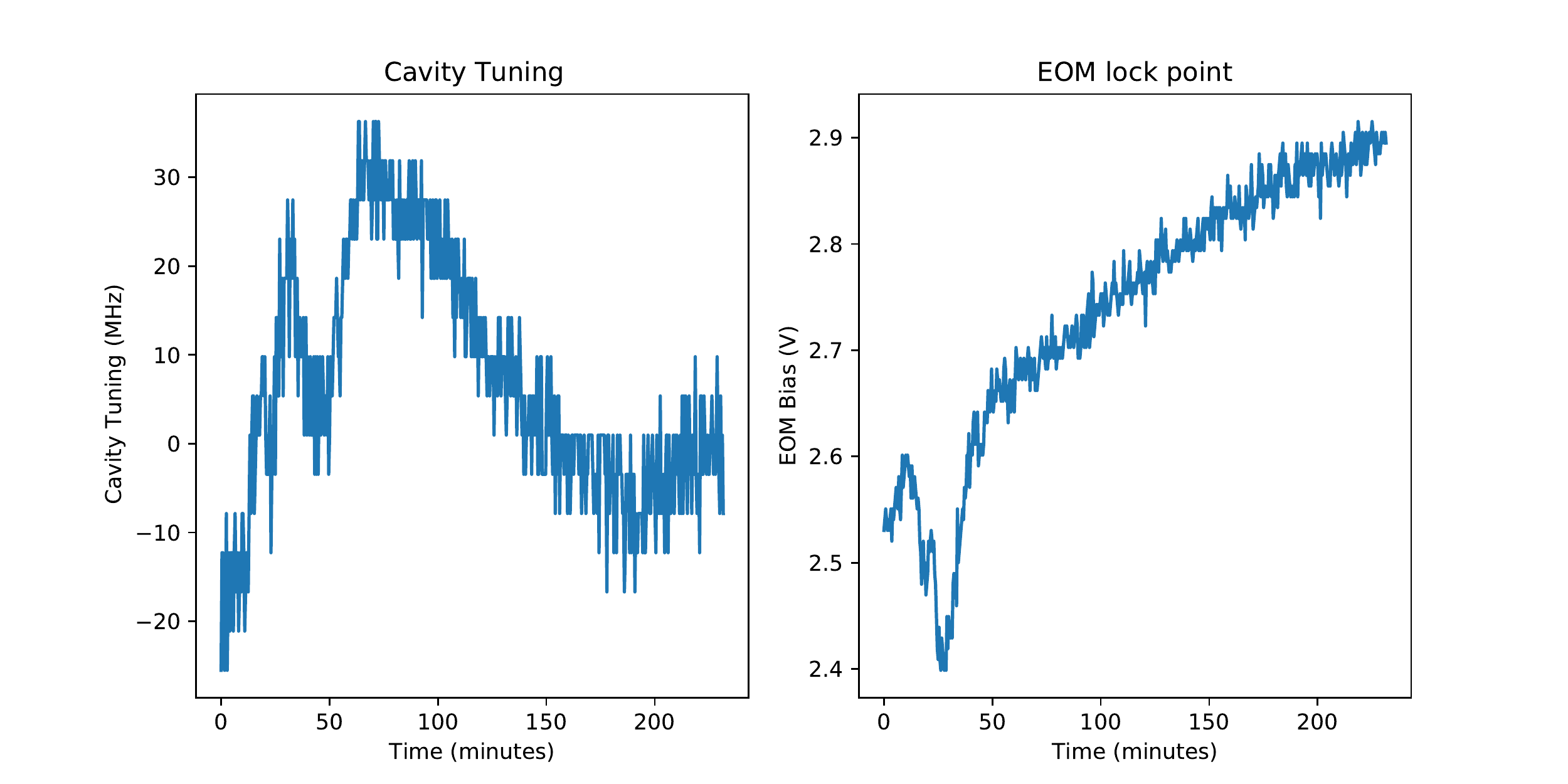}
  \caption{Drift of the EOM lock bias and cavity during a 4 hour experiment. For reference the cavity FWHM is 238 MHz and the EOM $V_\pi$ is 1.4 volts.}
  \label{fig:lock}
\end{center}
\end{figure*}

The free spectral range of the cavity is 75.11 GHz and the full width half maximum linewdith (FWHM) is 238 MHz, consistent with the mirror properties. We apply voltages amplified by a Thorlabs MDT693B to the piezo to tune the cavity at 1.83 GHz/V. After the filter cavity, the light is coupled back into a fiber by a symmetric set of lenses and collimators and sent directly to the heralding SNSPD. The end-to-end efficiency of the filter cavity is measured to be 9\% . The filter cavity resonance is relatively stable over the course of the experiment, while the optimal bias for suppressing the carrier in the amplitude modulator EOM can drift up to a quarter of $V_{\pi}$ (Fig \ref{fig:lock}).

\section{Experimental Sequence}

Our experimental sequence for entanglement consists of two parts, a presequence to check the state of the SiVs, and the entanglement sequence proper. We find that SiV A occasionally ionizes, with the spectral feature disappearing entirely, and can be revived by applying light at 532~nm. In contrast, the spectral feature for SiV B never disappears, but does hop between several metastable frequencies. The diffusion of SiV B happens faster when light at 737~nm is applied.

To check the state of the SiVs, we start by attempting to initialize them. For the initialization sequence, we apply a laser (ECDL for SiV A and TiSaph for SiV B) for 10us and apply a $\pi$ pulse if recorded counts are higher than 7 in the case of SiV B or lower than 7 in the case of SiV A.

If initialization fails for SiV A, we apply green light for 1~ms. If initialization for SiV B fails, we apply 737~nm light for 1~ms. We then proceed to the main experimental sequence. This procedure does not deterministically put the SiVs in the correct state, so we also postselect entanglement events where most of the surrounding unheralded trials ended with the correct readout.

Following our initialization sequence, we perform 200 trials of the main entanglement experiment. In each trial, we take the readout from the preceding trial as the measurement for conditional initialization. This is followed by the heralded entanglement sequence and finally the read out of the spin states. 

\section{Echo Sequence and MW Characterization}

\begin{figure*}
\begin{center}	
  \includegraphics[width=\textwidth]{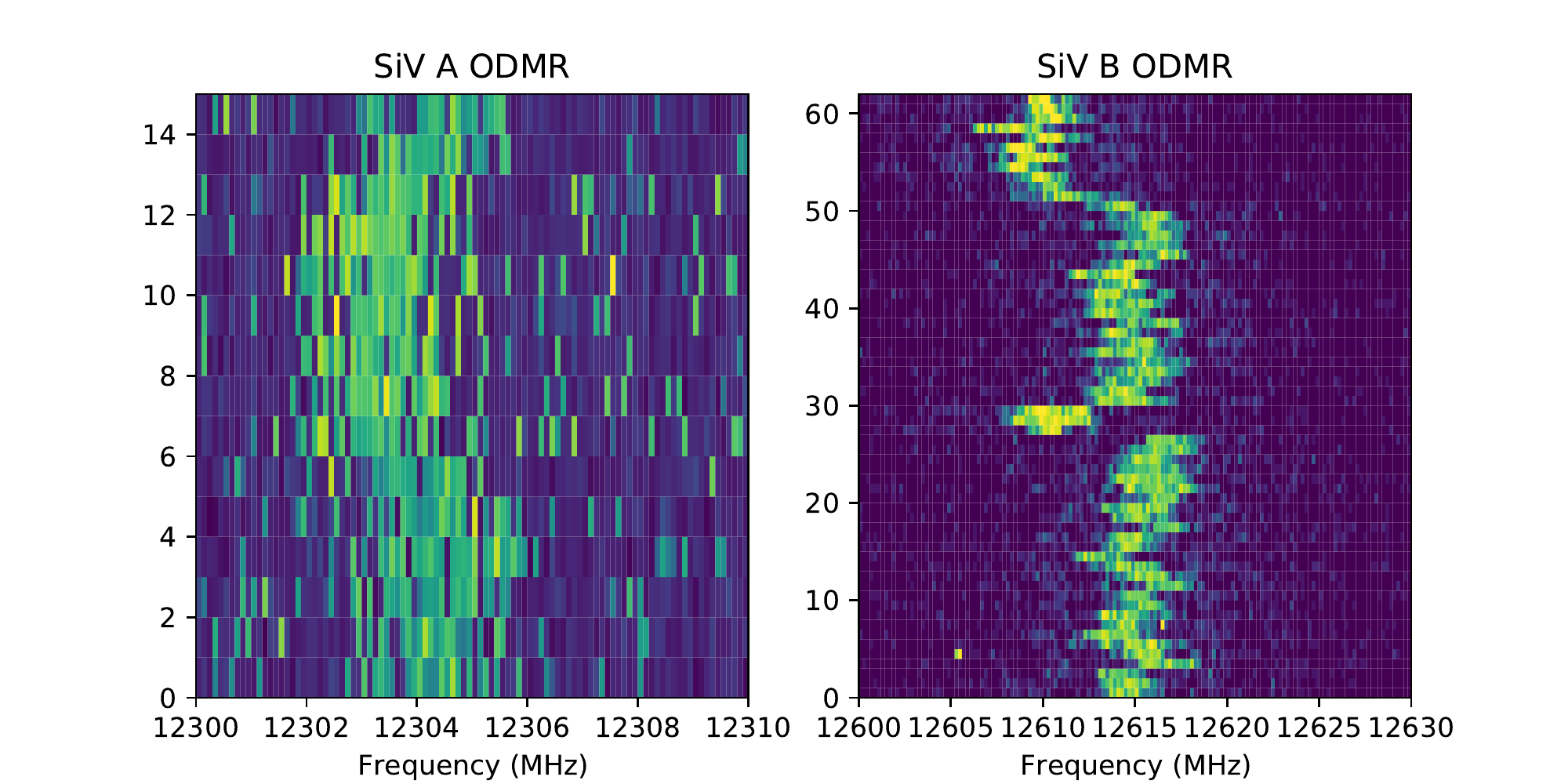}
  \caption{Optically detected magnetic resonance measurements on SiV A and SiV B. SiV A exhibits slow variation in qubit frequency.}
  \label{fig:odmr_drift}
\end{center}
\end{figure*}

\begin{figure*}
\begin{center}	

  \includegraphics[width=\textwidth]{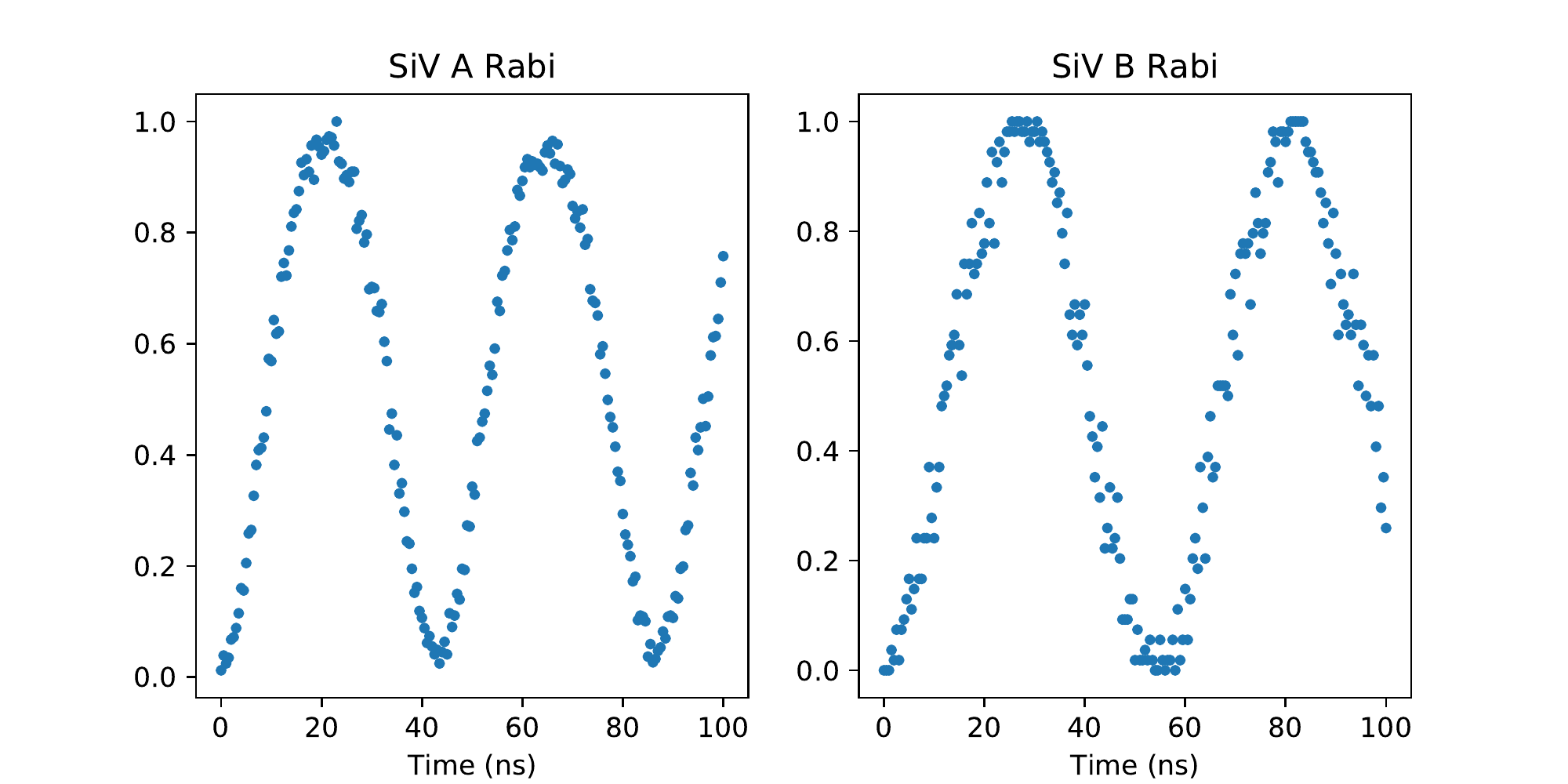}
  \caption{Rabi driving on SiV A and SiV B.}
  \label{fig:rabi}
\end{center}
\end{figure*}

The heralding pulses are applied in the middle of a series of microwave pulses that decouple the qubits from noise and put them in an equal superposition of two-qubit states. First we apply a $\pi/2$ rotation around the Y-axis to SiV A and then SiV B sequentially (11 and 14 ns). We then wait 401 ns. During this window we apply a 200 ns optical probe pulse, starting roughly 100 ns after the $\pi/2$ rotations Subsequently, we apply a $\pi$ rotation around the X-axis on SiV A and then SiV B (22 and 28 ns, Fig \ref{fig:rabi}). We then wait 387 ns and apply a $\pi/2$ rotation on SiV A around the negative Y-axis. This completes a Hahn echo on SiV A with $\tau = 426$ ns. Finally we wait 25 ns and apply a $\pi/2$ rotation on SiV B around the negative Y-axis, which completes a Hahn echo on SiV B with $\tau = 437$ ns. To read out in the Y basis, we do the final $\pi/2$ rotations around the X-axis instead and to read out in the Z basis we omit the final rotations entirely.

\section{$T_2$ Time}

\begin{figure*}
\begin{center}	

  \includegraphics[width=.7\textwidth]{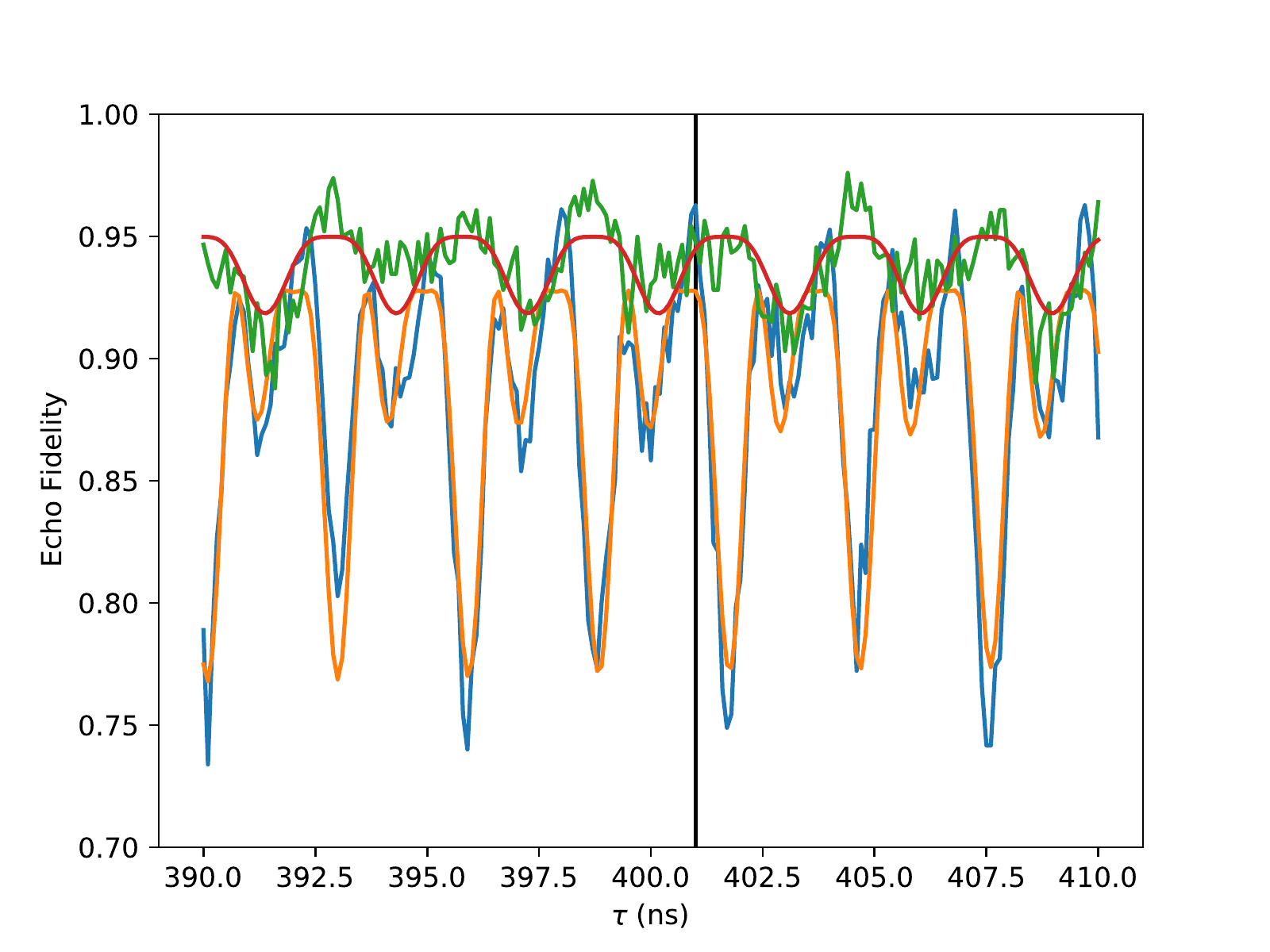}
  \caption{Results of a simultaneous Hahn Echo performed on SiV A and SiV B. Pulses are applied consecutively. Vertical line indicates value of $\tau$ used for the entanglement experiment.}
  \label{fig:T2both}
\end{center}
\end{figure*}

To characterize the fidelity of our spin system, we perform the simultaneous Hahn echo sequence on our spins, in the absence of any probe pulses (Fig \ref{fig:T2both}). Both echo curves have collapses and revivals with the echo curve on SiV A revealing a beating pattern between two frequencies. These frequencies (342 MHz and 686 MHz for SiV A) are considerably higher than expected Larmour frequencies of nuclear spins in our field ($\approx$5~kG) while being much lower than the expected Larmour frequencies of other SiV spins. However, these beats could be aliased by the Nyquist frequency of our sweep (10 GHz) which would mean the true frequencies are 10.342 GHz and 10.686 GHz. These frequencies are plausible for an SiV electron spin with moderate hyperfine interaction with SiV A. 


We choose $\tau = 401$ ns to minimize coupling to this spin. At this value, the average fidelity is 0.96 for SiV A and 0.95 for SiV B. However, over the course of the experiment, we saw diffusion of the SiV A qubit frequency (Fig \ref{fig:odmr_drift}), on a timescale suggestive of changes in gyromagnetic ratio caused by strain, rather than magnetic fluctuations of a $^{13}C$ bath. These gyromagnetic ratio changes would have changed the hyperfine interaction with the closely coupled spin. Decoherence caused by changes in the hyperfine coupling could explain both the fluctuation in observed fidelity of the echo sequence during the entanglement experiment, as well as its overall smaller value. 

\section{Bounds on Magnetic Dipole Interaction}

\begin{figure*}
\begin{center}	

  \includegraphics[width=\textwidth]{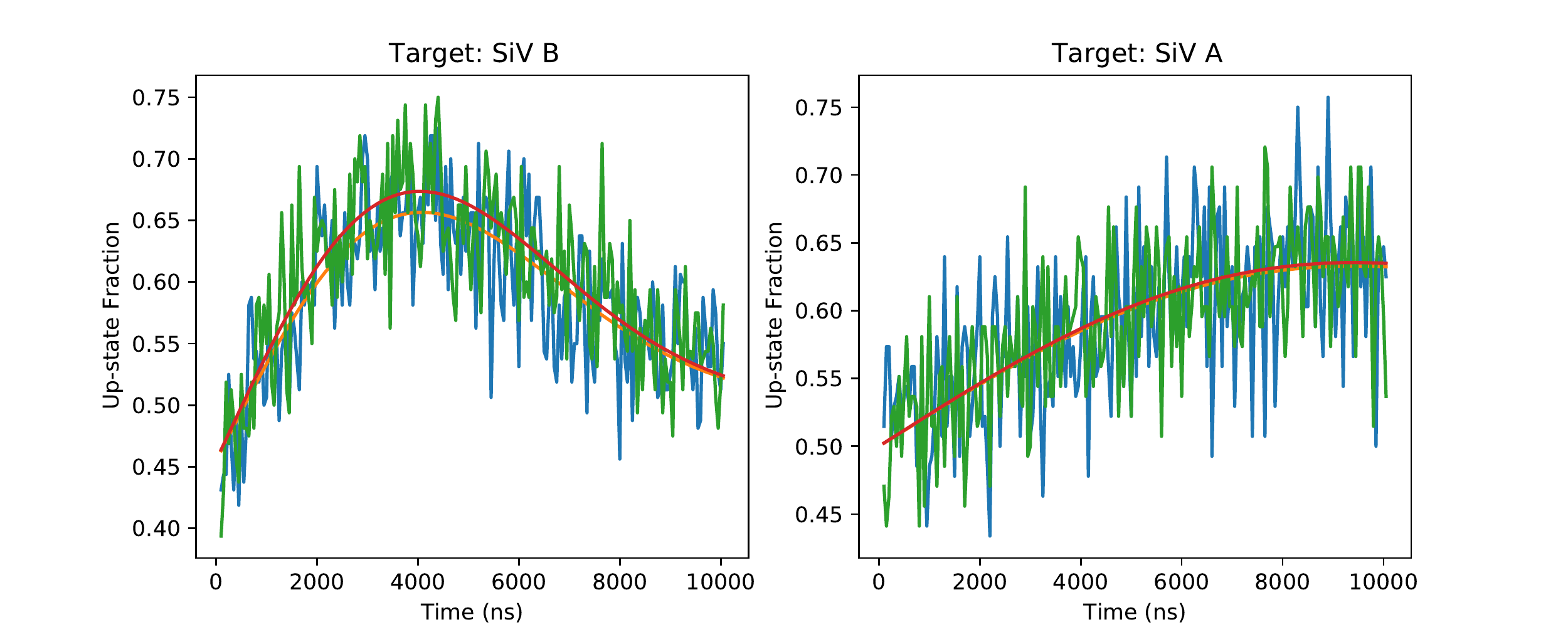}
  \caption{Results of a DEER sequences performed between SiV A and B. Blue data and orange fit are with control spin up and green data and red fit are with control spin initialized down. Fits are a sinusoid multiplied by a decaying $T_2$ envelope. Common mode sinusoid behavior is from drive-qubit detuning.}
  \label{fig:DEER}
\end{center}
\end{figure*}

Given the strongly coupled spins implied by the dual Hahn Echo sequence, we performed experiments to rule out magnetic dipole interactions between SiV A and B. We used a dual electron-electron resonance (DEER) sequence to isolate the coupling between the SiVs from any environmental effects. 

The DEER sequence consists of a Hahn Echo sequence applied to the target spin where the last $\pi/2$ pulse has a Y rather than X phase and a $\pi$ pulse applied to the control spin in the middle of the echo sequence. Initializing the control spin in the up vs down state before the sequence will produce oscillating fields with different signs which the target spin will be sensitive to in the event of nonzero magnetic coupling.

We run DEER sequences with SiV A and B as both target and control spin (Fig \ref{fig:DEER}), and find the difference between the results with the control spin up vs down implies a coupling of at most $6\pm2 \times 10^2$ Hz between the two SiVs, which would produce negligible interaction during our 800 ns long entanglement sequence.

\section{Coupling Efficiency Calibration}

\begin{figure*}
\begin{center}	

  \includegraphics[width=.7\textwidth]{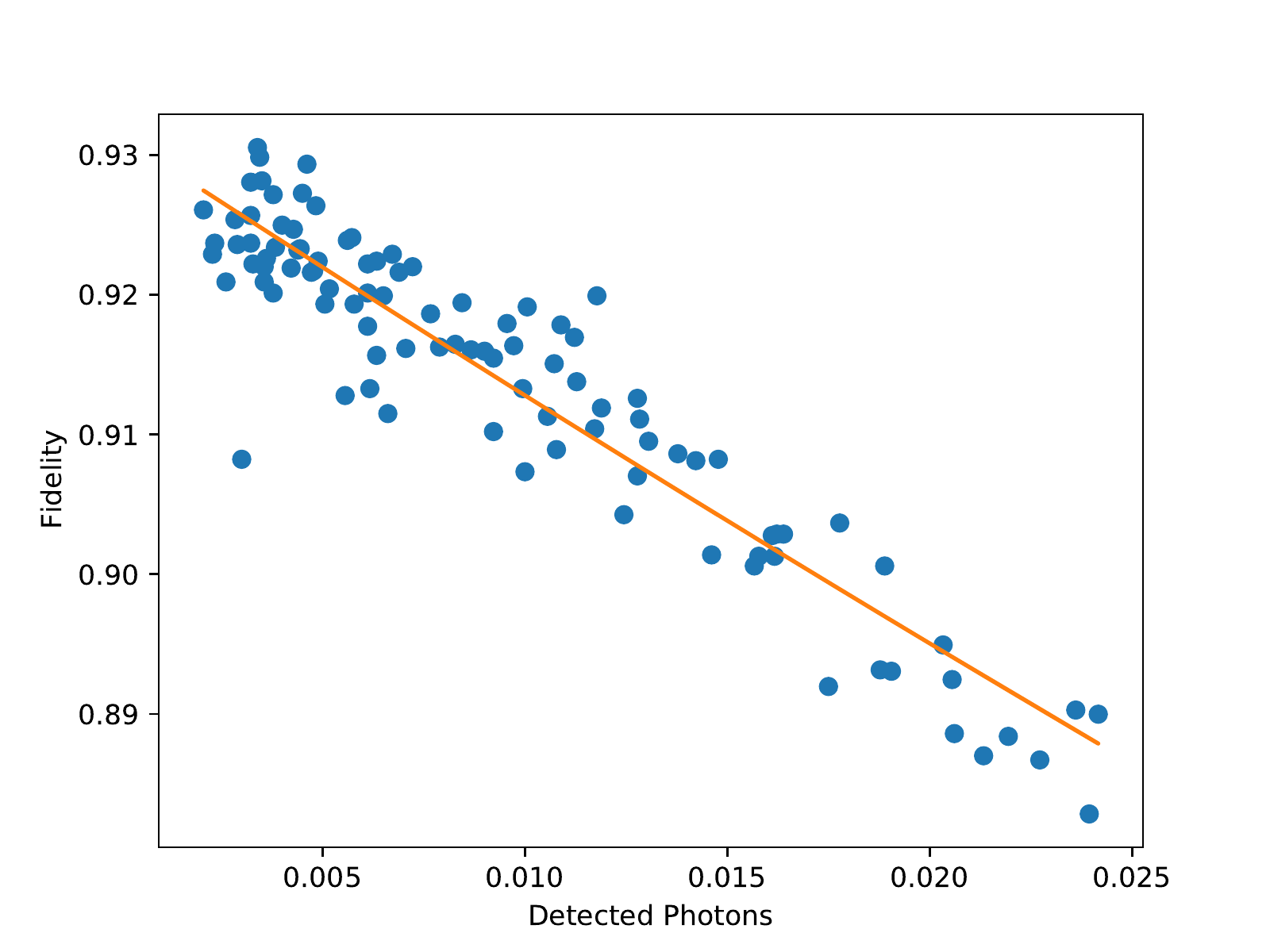}
  \caption{Fidelity of a Hahn Echo on SiV  B vs the number of photons detected by the readout SNSPD}
  \label{fig:coupling}
\end{center}
\end{figure*}

We characterize our fiber-device coupling efficiency with two independent experiments. First, we apply broadband light from a supercontinuum laser through the fiber and measure the reflected intensity. By comparing the reflection to a calibration value obtained with a retro-reflector spliced to the end of the fiber, we calculate an 86\% fiber-device coupling efficiency. \cite{Bhaskar2019}

In addition, we perform an experiment we we reflect light from our device at the frequency of greatest contrast for SiV B during one window of a Hahn Echo sequence on SiV B. Every photon that scatter from the device has a 50\% chance of flipping the electron to the wrong state. Roughly half of the photons incident on the device are reflected back. Of those, a fraction denoted by $\eta_{wg}$ make it to the 50/50 beamsplitter. Extrapolating from the decrease in fidelity per detected photon (Fig \ref{fig:coupling}), we estimate $\eta_{wg} = 84 \pm3\%$ using this method.

We measure the transmission of our filter cavity $\eta_{cav} = 0.09$ with photodiode measurements during cavity alignment and by comparing the counts on the readout and heralding SNSPDs when light at the filter cavity frequency is reflected off of the nanocavity.

At the drive power we operate at, 31\% of incident power at the phase EOM is transferred to each sideband. Constructive interference doubles this power in the interferometer experiment. Given our heralding rate of $6\cdot{}10^{-4}$ per experiment, and the fact that we prepare the two SiVs in an equal mixture of all four Bell states, we calculate that the mean photon number at the cavity is $\langle n\rangle =0.106$, which translates to a 5.3\% decoherence due to two-photon events.

\section{Cooperativity}

\begin{figure*}
\begin{center}	

  \includegraphics[width=.7\textwidth]{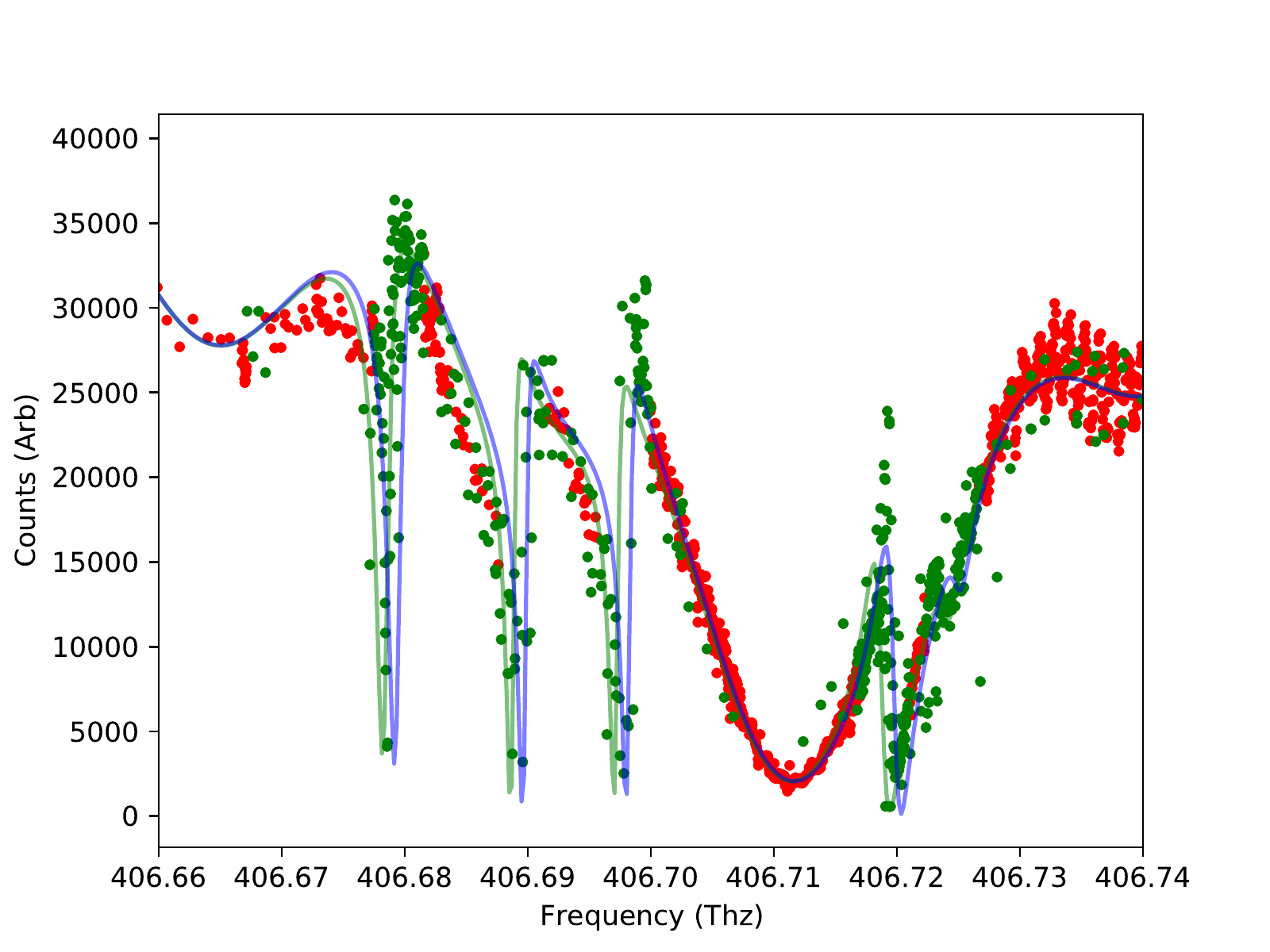}
  \caption{ECDL scan of the SiV-cavity system (red, dark green) and fit for the $\left|\downarrow\right\rangle$ (solid blue) respectively $\left|\uparrow\right\rangle$ state (solid light green), assuming a sinusoidal wavelength-dependent modulation of laser intensity. Points near SiVs are excluded, because SiVs are not initialized for this scan, as well as outliers (dark green).}
  \label{fig:SiVscanbig}
\end{center}
\end{figure*}

To obtain the cavity parameters, as well as the coupling strength of the SiVs to the cavity field, we perform several laser scans of the cQED system. We initially scan the laser frequency across the the cavity without initializing the SiVs. This data is fitted with a model including several cavity coupled SiVs and a sinusoidal modulation of the input power: The amplitude reflectivity of this system is 
\begin{equation}
    R(\omega_l, \omega_{k}) = 1 - \frac{2\kappa_w}{i\Delta_c + \kappa_{tot} + \sum_k \frac{g_k^2}{i\Delta_{a,k}+\gamma_{k}}},
    \label{eq:cav_amplitude}
\end{equation}
where $\Delta_{c} = \omega_l - \omega_{c}$ represent the detuning of the laser (at frequency $\omega_l$) from cavity resonance (at frequency $\omega_c$). Similarly, for each SiV, labeled by index $k$, $\Delta_{a,k} = \omega_l - \omega_{k}$ is the detuning of the laser from the resonance of SiV $k$ (at frequency $\omega_{k}$). The total cavity loss rate $\kappa_{tot} = \kappa_w + \kappa_l$ is the sum of the scattering rate $\kappa_l$ and leakage into the waveguide ($\kappa_w$).
After a rough fit of all parameters, the data points close to the SiVs are excluded, as quantum jumps of the spin states result in noisy data close to their resonances (see Fig \ref{fig:SiVscanbig}). The remaining spectrum is fitted with an additional weight on the center of the cavity, thereby ensuring that the ratio of loss to waveguide coupling $\kappa_l/\kappa_w$ is correct, as this has the strongest influence on the final model. We know that $\kappa_l < \kappa_w$ due to the fact that several SiV resonance dip below the lowest point of the cavity resonance.

\begin{figure*}
\begin{center}	

  \includegraphics[width=.7\textwidth]{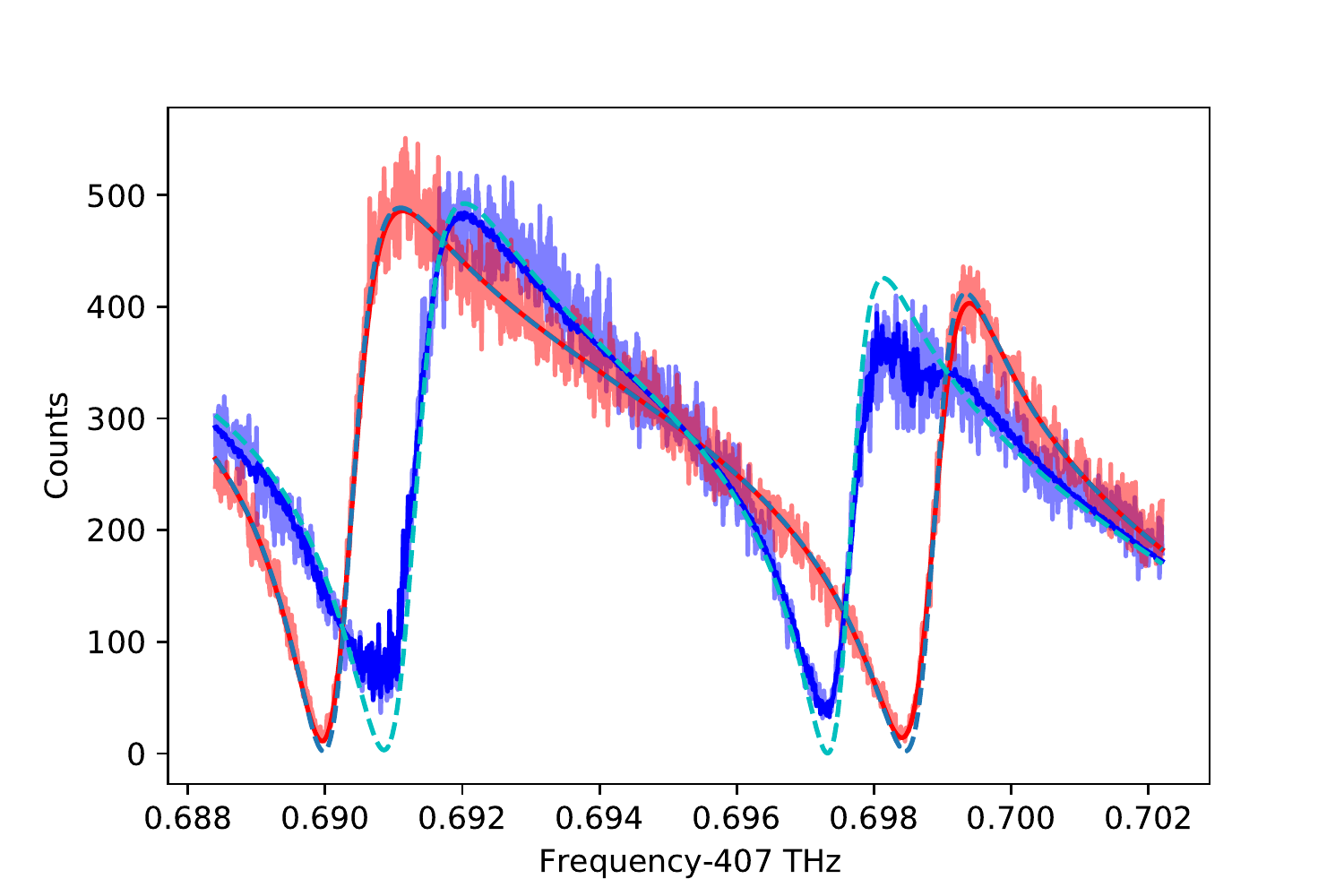}
  \caption{TiSaph scan of the SiV-cavity system ($\left|\uparrow\downarrow\right\rangle$: light red, $\left|\downarrow\uparrow\right\rangle$: light blue) and fit taking into account spectral diffusion and state of SiV after initialization attempt ($\left|\uparrow\downarrow\right\rangle$: dark red, $\left|\downarrow\uparrow\right\rangle$: dark blue). Dashed lines: expected spectrum with ideal initialization and no optical diffusion. See also Fig. \ref{fig:SiVscanzoom}.}
  \label{fig:SiVscansmall}
\end{center}
\end{figure*}

\begin{figure*}
\begin{center}	

  \includegraphics[width=.7\textwidth]{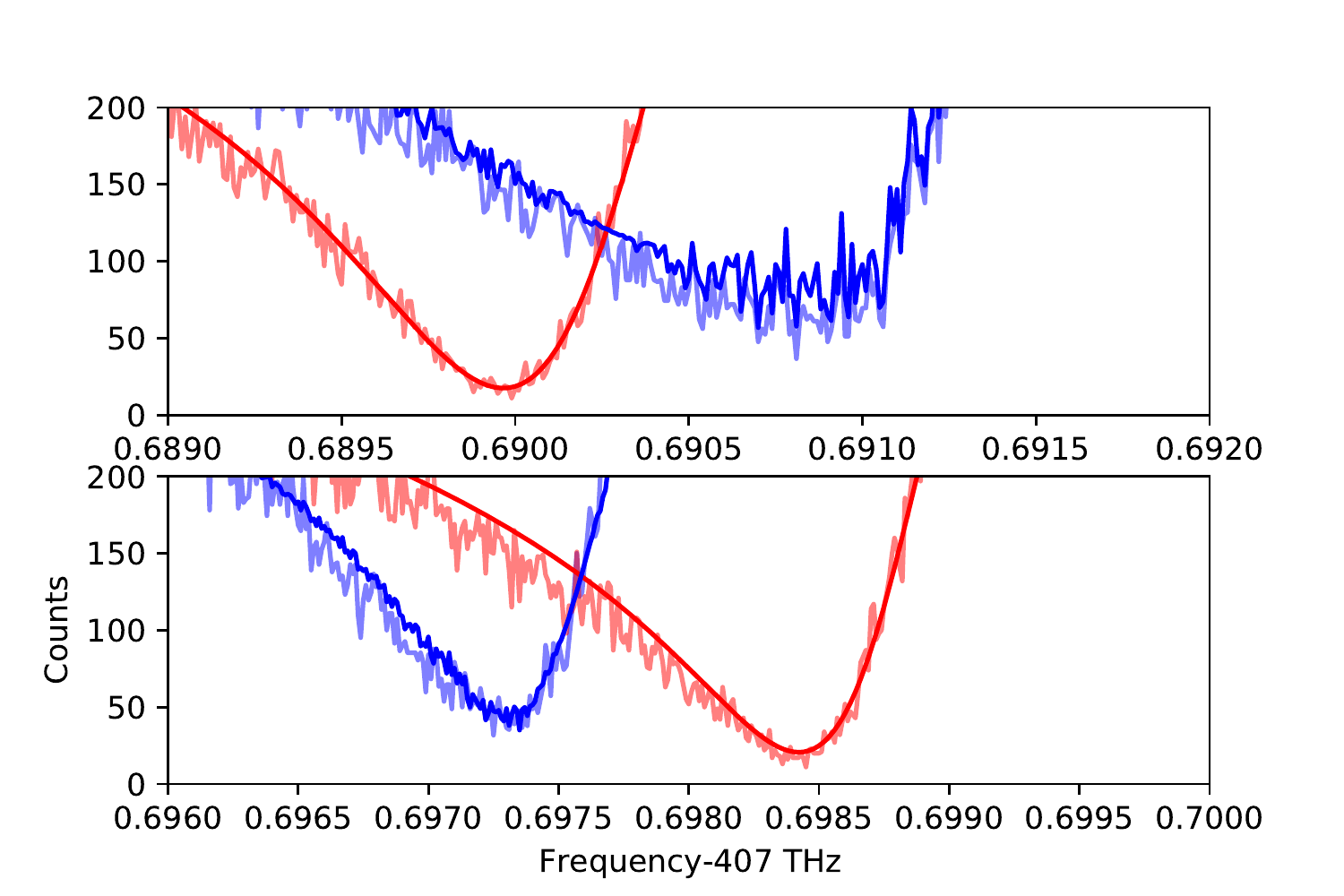}
  \caption{Detail of fits of TiSaph scan of the SiV-cavity system in \ref{fig:SiVscansmall}. Precise matching of the dips in the spectrum is important, as the residual reflection off the $\left|\downarrow\right\rangle$ states dominates the interference pattern in the phase scan of the interferometer (main text Fig. 3a). Noise on the fit of the $\left|\downarrow\uparrow\right\rangle$ state (dark blue) is a result of the imperfect initialization fidelity measured at each laser detuning. Contrast of the $\left|\uparrow\downarrow\right\rangle$ state (red) appears superior due to better initialization.}
  \label{fig:SiVscanzoom}
\end{center}
\end{figure*}

We then proceed with a narrow scan across the frequencies of SiV A and B, initializing them in the $\left|\uparrow\downarrow\right\rangle$ respectively $\left|\downarrow\uparrow\right\rangle$ state. 
By measuring the spin state again after the laser scan we furthermore extract the initialization fidelity of each spin during the laser scan. We use the same model \ref{eq:cav_amplitude} to fit the spectra (Fig \ref{fig:SiVscansmall}, Fig \ref{fig:SiVscanzoom}), this time adding optical diffusion of the optical transition frequency of the SiVs  with a Gaussian probability distribution, and taking into account the uncertainty in the data due to shot noise. The relevant cQED parameters from this fit are detailed in Tab. \ref{tab:cavityparameters}. We find that the resulting cooperativity 
$$ C_k = \frac{g_k^2}{\kappa_{tot} \gamma_k}$$
for both SiVs is robust against possible sources of systematic uncertainty, such as offset in the broad spectrum (Fig \ref{fig:SiVscanbig}) beyond the calibrated background counts.

\setlength{\belowcaptionskip}{0pt}
\begin{table}

\begin{ruledtabular}
\begin{tabular}{lccc}
Parameter & SiV A & SiV B & Cavity\\
\hline
resonance & $\omega_{A,\uparrow}=2\pi\cdot406.692$ THz & $\omega_{B,\uparrow}=2\pi\cdot406.699$ THz & $\omega_c=2\pi\cdot406.706$ THz\\ \hline
detuning  & $\Delta_A=2\pi\cdot14.6$ GHz & $\Delta_B=2\pi\cdot7.2$ GHz & \\ \hline
optical line splitting & $\omega_{A,\uparrow}-\omega_{B,\downarrow}=2\pi\cdot0.95$ GHz &  $\omega_{B,\uparrow}-\omega_{B,\downarrow}=2\pi\cdot1.23$ GHz & \\ \hline
natural linewidth & $\gamma_A=2\pi\cdot80$ MHz & $\gamma_A=2\pi\cdot97$ MHz & $\kappa_{tot} = 2\pi\cdot14.5$ GHz\\
&&&$\kappa_{w} = 2\pi\cdot9.0$ GHz\\
&&&$\kappa_{l} = 2\pi\cdot5.4$ GHz\\ \hline
spectral diffusion & $\sigma_A=2\pi\cdot58$ MHz & $\sigma_B=2\pi\cdot113$ MHz  & \\ \hline
coupling strength & $g_A=2\pi\cdot4.1$ GHz & $g_B=2\pi\cdot2.9$ GHz\\ \hline
cooperativity (instantaneous) & $C_A = 6.1$ & $C_B=14.4$&\\ \hline
cooperativity (averaged) & $\bar C_A = 4.0$ & $\bar C_B=11.7$&
\end{tabular}
\end{ruledtabular}
\caption{\label{tab:cavityparameters} Relevant parameters of the cavity QED system. 
}
\end{table}

\section{Readout Fidelity and Cyclicity}

\begin{figure*}
\begin{center}	

  \includegraphics[width=\textwidth]{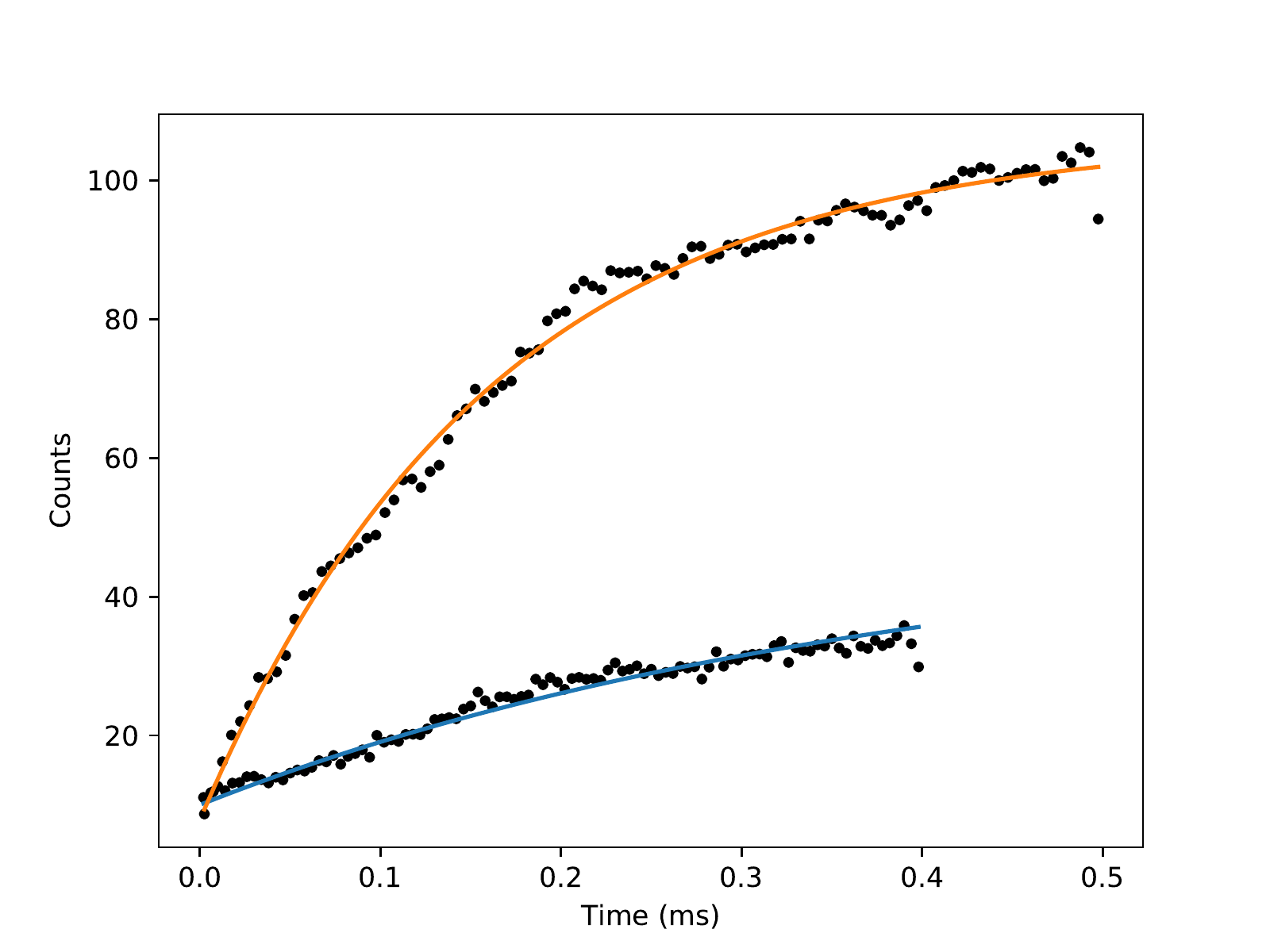}
  \caption{Average counts under laser illumination vs time for SiV A (orange) and B (blue) after being initialized in the spin down (non-reflective) state. }
  \label{fig:cyclicity}
\end{center}
\end{figure*}

To find the branching ratios for our optical transitions, we initialize SiV A or B in the non-reflecting state and then apply the readout laser until the averaged reflected counts converge to half way in between the up and down state reflectivities (Fig \ref{fig:cyclicity}). We use the time constant of this curve, plus the measured count rate in the reflective state at the same laser power  to calculate the mean number of reflected photons necessary to flip each spin. We find $r_a =  3.8 \pm 0.1 \times 10^3 $ and $r_b =  9.2 \pm 1.3 \times 10^3 $.

\begin{figure*}
\begin{center}	

  \includegraphics[width=\textwidth]{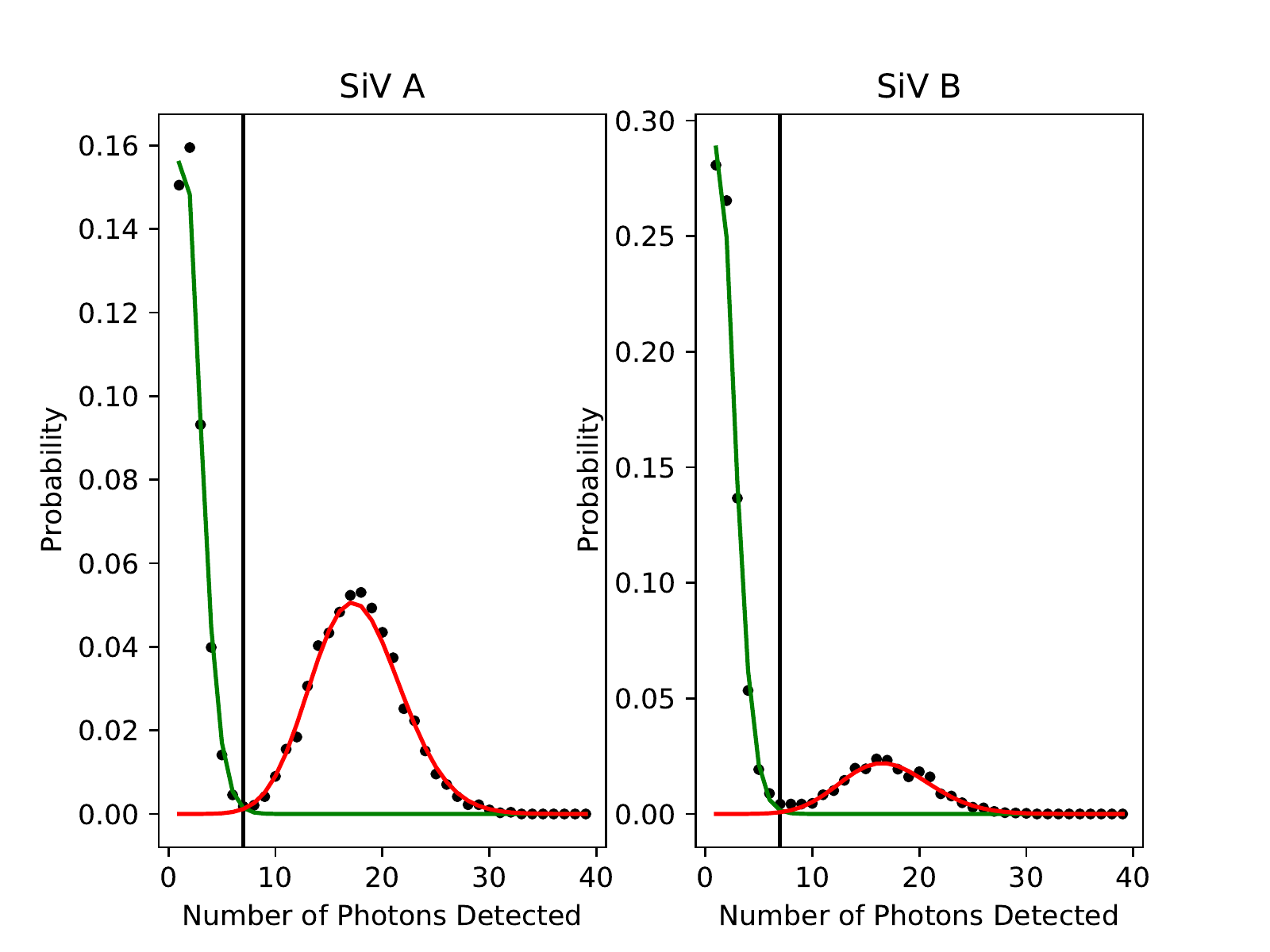}
  \caption{The histograms of photons received after a 10us readout pulse following a pi-half rotation on each SiV, fit with two Poissonian distributions with mean values 1.9 and 17.7 for SiV A low and high and 1.7 and 17.0 for SiV B low and high, respectively.}
  \label{fig:readout_stats}
\end{center}
\end{figure*}

In addition, we apply lasers for a 10us window to each SiV  and look at the histogram of received counts (Fig \ref{fig:readout_stats}). We then fit the histogram with two Poisson distributions and calculate the probability of the up state being misidentified as the down state or vice versa. We find that the odds of correctly identifying the state are 99.89\% for SiV A and 99.87\% for SiV B.

We evaluate our readout and initialization fidelity by initializing both spins in the down state approximately 30,000 times and then reading the state of both spins. We find that the fidelity $F_{rA} = 99.839\pm 0.006\%$ and $F_{rB} = 99.910\pm0.0045\%$ observed during these runs is in good agreement with the fidelity predicted by the branching ratio and readout histograms.

\section{The Role of SiV-cavity Detuning in Contrast}

Our CQED system is able to achieve high spin-dependent contrast for a range of atom-cavity detunings with only moderate cooperativities. To see why, we consider the reflected field from the cavity given by:
$$ R = 1 - \frac{2\kappa_w}{i\Delta_c + \kappa_{tot} + \frac{g^2}{i\Delta_a+\gamma}},$$
where $\Delta_{a,c} = \omega - \omega_{a,c}$ represent the detuning from the SiV and caviy, respectively and the total cavity loss rate is a product of the scattering loss and leakage into the waveguide $\kappa = \kappa_w + \kappa_l$.

For the SiV feature to dip to zero reflectively, the denominator must meet the condition
$$i\Delta_c + \kappa_{tot} + \frac{g^2}{i\Delta_a+\gamma} = 2\kappa_w,$$
which is satisfied when 
$$\Delta_a = \sqrt{\frac{g^2\gamma}{2\kappa_w - \kappa} - \gamma^2}$$
and
$$\Delta_c = \frac{2\kappa_w-\kappa}{\gamma}\sqrt{\frac{g^2\gamma}{2\kappa_w - \kappa} - \gamma^2}.$$
This implies that the optimal SiV-cavity detuning is

$$ \Delta = \Delta_c-\Delta_a = \left(\frac{2\kappa_w - \kappa}{\gamma} - 1 \right)\sqrt{\frac{g^2\gamma}{2\kappa_w - \kappa} - \gamma^2}.$$
Assuming that $\gamma<<g,\kappa_w-\kappa_l$:
$$\Delta = g\sqrt{\frac{\kappa_w - \kappa_l}{\gamma}}$$

To see how sensitive the contrast is to changes in the SiV-cavity tuning, we can hold $\Delta_a$ fixed and let $\Delta_c$ vary.
Then for small deviations ($\delta_c$) around the detuning that would lead to a reflection dip to zero, we have:
$$R(\delta_c) = 1 - \frac{2\kappa_w}{2\kappa_w + i\delta_c}$$
If the detuning $\Delta_c \geq \kappa$, then the reflection near the SiV dip will approach 1. Thus, to maintain 1:10 contrast of the feature $|\delta_c| \leq \frac{2}{3}\kappa_w$. In practice, this means that we can maintain 90\% spin-dependent contrast for a given SiV over a range of atom-cavity detunings comparable to the cavity linewidth.

\section{Theoretical model of the system}


We model our system using the cavity QED parameters described in Tab. \ref{tab:cavityparameters}. The systems is probed by the two laser sidebands at the frequencies $\omega_{laser,A}=406.6910$~GHz and $\omega_{laser,B}=406.6984$~GHz. 
The amplitude of the photonic wavefunction in transmission of the cavity is 
\begin{equation}
    T(\omega_A, \omega_B) \propto c_\textrm{carrier}R(\omega_\textrm{carrier}, \omega_A, \omega_B) e^{i\phi_c} + c_\textrm{sideband} R(\omega_\textrm{sbA}, \omega_A, \omega_B) e^{-i\Delta\phi/2} + c_\textrm{sideband} R(\omega_\textrm{sbB}, \omega_A, \omega_B) e^{i\Delta\phi/2},
\end{equation}
with the cavity reflection coefficient $R$ defined in equation \eqref{eq:cav_amplitude}, the carrier frequency $\omega_{carrier} = (\omega_{sbA}+\omega_{sbB})/2$, the sideband frequencies $\omega_{sbA}$ and $\omega_{sbB}$, and the spin-dependent SiV resonances $\omega_A\in\{ \omega_{A, \uparrow}, \omega_{A, \downarrow}\}$ and $\omega_B\in\{ \omega_{B, \uparrow}, \omega_{B, \downarrow}\}$. The coefficients for the carrier leakage $c_\textrm{carrier}=0.08$ and the sideband efficiency $c_\textrm{sideband}=0.38$ are experimentally determined by characterizing the EOMs. The phase of the carrier $\phi_c$ is in the ideal case close to 0, but can vary due to dispersion and drifts in the lock point of the intensity modulator EOM$_\textrm{MZ}$, with values possible in the range of $\phi_c\in [-\pi/2,\pi/2]$. For the phase scan, we set $\phi_c=0$, as EOM$_\textrm{MZ}$ was locked stringently, whereas we leave it as a free parameter for the correlation experiments, where it was locked only occasionally. While the carrier phase only has a minor influence on the predicted fidelity, the YY correlations and the XX correlations, strongly influences the balance between the $\left|\uparrow\downarrow\right\rangle_\textrm{\!\tiny{AB}}\!$ and $\left|\downarrow\uparrow\right\rangle_\textrm{\!\tiny{AB}}\!$ state. We thus choose the carrier phase which matches this balance, finding $\phi_c=0.398\pi$.

The relative phase of the sidebands $\Delta\phi=2\phi_\mu$ is controlled by the relative phase of the microwave signals $\phi_\mu$ driving the two EOMs. Instead of directly applying a phase shift, we make use of the differential (optical) path length $\Delta L$ between the optical and the microwave path and control the relative phase $\phi_\mu=\Omega \Delta L/c$ by fine tuning the microwave frequency $\Omega$, with the speed of light $c$. As the free spectral range of the microwave-optical interferometer $c/\Delta L\sim 4.7$~MHz is far below the size of the spectral features, this does not adversely effect the performance of the protocol. 

A spin state $\left|\psi_{in}\right\rangle_\textrm{\!\tiny{AB}}\! \propto \alpha_{\uparrow\uparrow} \left|\uparrow\uparrow\right\rangle_\textrm{\!\tiny{AB}}\! +\alpha_{\uparrow\downarrow} \left|\uparrow\downarrow\right\rangle_\textrm{\!\tiny{AB}}\! +\alpha_{\downarrow\uparrow} \left|\downarrow\uparrow\right\rangle_\textrm{\!\tiny{AB}}\! +\alpha_{\downarrow\downarrow} \left|\downarrow\downarrow\right\rangle_\textrm{\!\tiny{AB}}\! $ upon detection of a heralding photon in transmission of the cavity is projected into 
\begin{equation}
    |\psi_{out}\rangle \propto T(\omega_{A, \uparrow}, \omega_{B, \uparrow}) \alpha_{\uparrow\uparrow} \left|\uparrow\uparrow\right\rangle_\textrm{\!\tiny{AB}}\! + T(\omega_{A, \uparrow}, \omega_{B, \downarrow})\alpha_{\uparrow\downarrow} \left|\uparrow\downarrow\right\rangle_\textrm{\!\tiny{AB}}\! + T(\omega_{A, \downarrow}, \omega_{B, \uparrow}) \alpha_{\downarrow\uparrow} \left|\downarrow\uparrow\right\rangle_\textrm{\!\tiny{AB}}\! + T(\omega_{A, \downarrow}, \omega_{B, \downarrow}) \alpha_{\downarrow\downarrow} \left|\downarrow\downarrow\right\rangle_\textrm{\!\tiny{AB}}\!
\end{equation}

To complete the model, we need to properly account for non-ideal local qubit operations, resulting from non-ideal pulse parameters, off-resonant driving and decoherence.

We extract these parameters form spin measurements performed during the correlation experiment when no heralding photon was detected. As the predicted fidelity is most sensitive to the decoherence, we alternate YY and ZZ basis measurements with an XX basis measurement, which in the unheralded case is a simple Hahn-Echo experiment measuring the decoherence of the spins during the sequence. This furthermore allows for post-selecting the heralded data based on the fidelity of the $N=500$ XX measurements closest to the heralding event to exclude instances when an SiV was ionized, or the Zeeman splitting $Z_A$ or $Z_B$ was far off the assumed value. Furthermore, this ensured that SiV B was effectively decoupled from the dark spin observed in Fig. \ref{fig:T2both}. We include data with a Hahn Echo fidelity better than $1-\mathcal{F}_\textrm{HE,A}\leq 0.17$ and $1-\mathcal{F}_\textrm{HE,B}\leq 0.15$. We further exclude extreme outliers in the $N=500$ YY or ZZ measurements closes to the heralding event. This only occurs very rarely and does not significantly influence the fidelity measurement but improves the convergence of the fitted microwave errors, improving the predictive power of our model. 

While fitting the ZZ distributions allows for constructing a realistic model of the pulse amplitude, detuning and spectral diffusion of the Zeeman splitting, there remain two undetermined parameters, the angle error of the readout basis, and the relative phase of the two microwave tones. The angle error originates from jitters in the AWG timing and can be estimated from the unheralded YY correlation data to be on average of the order of $0.5$ to $0.10$ radians. However, we cannot accurately determine a variance of the angle error or infer the angle error of the X axis measurements, so that we do not include this in the model. This can lead to an overestimation of the decoherence from the unheralded XX measurements, consistent with heralded correlation data in the XX axis. The overestimation of the decoherence from the average Y angle error corresponds to an increase in expected fidelity by $0.003$, which is treated as a systematic uncertainty.

The relative phase between the pulses addressing SiV A and SiV B is well defined at the source, but can be different at the device due to dispersion in the microwave cables to the sample. This is exacerbated by resonances in the cryogenic coaxial cables observed in our system. In our system, the microwave dispersion was not measured at cryogenic temperatures, so we have no reliable information about the actual relative phase of the two frequencies at the sample. We therefore simulate the system for a set of 24 equally distributed microwave phases between 0 and $2\pi$. The cited predicted values of the model are averaged over all microwave phases, and the associated systematic uncertainty describes the standard deviation of the predicted values for all sampled microwave phases. 

This model is not uniquely defined due to the data it is based on and there are clear deviations between the measured data and the prediction by the model. These deviations however do not substantially influence the estimation of the fidelity

\section{Predicted Fidelity and Error Budget}

The model predicts fidelities in the range of $0.643\leq \mathcal{F}_{|\Psi^+\rangle}^\textrm{pred}\leq 0.695$ with an average value of $\left\langle\mathcal{F}_{|\Psi^+\rangle}^\textrm{pred}\right\rangle = 0.670$ and an uncertainty due to the angle of the read axis during the calibration of the spin decoherence of $^{+0.003}_{-0}$. This is consistent with the measured fidelity of $\mathcal{F}_{|\Psi^+\rangle}^\textrm{pred} = 0.710^{+0.019}_{-0.018}$. We note that the relative microwave phase resulting in the highest predicted fidelity, and therefore the best overlap with the measured value coincides with the highest consistency between the estimate of the read angle error of SiV A and SiV B.  

To understand the leading contributions the state preparation error, we eliminate the individual error sources from the model and compare the resulting fidelity with the predicted fidelity of the complete model. This marginal error contribution $\mathcal{E}_\textrm{source} = \mathcal{F}_{|\Psi^+\rangle}^\textrm{pred,source eliminated}-\mathcal{F}_{|\Psi^+\rangle}^\textrm{pred}$ for the dominant sources is provided in Tab. 1 of the main text as average value $\left\langle\mathcal{E}_\textrm{source}\right\rangle$ with the systematic error relating to uncertainties originating from the microwave dispersion, the exact probability of multi-photon states at the cavity, and the systematic uncertainty in the calibrated decoherence due to the estimated read angle error. 

Here, we add a few notes on the reason for some of the error sources. As evident from Fig. 2a and Fig. 3e in the main text, the residual reflection from the $\left|\downarrow\downarrow\right\rangle_\textrm{\!\tiny{AB}}\!$ state is a key contributor to the heralded state infidelity (note that the ZZ readout of in Fig. 3e is inverted due to the dynamical decoupling sequence). This primarily relates to the fact that SiV A was initialized in the $\left|\uparrow\right\rangle_\textrm{\!\tiny{A}}\!$ state, rendering the preselection of low optical spectral diffusion ineffective. As a consequence, the sideband frequency $\omega_\textrm{sbA}$ did not probe the spectrum at its maximum contrast point, leading to substantial residual reflection from the $\left|\downarrow\right\rangle_\textrm{\!\tiny{A}}\!$ state. This can in principle be compensated by adjusting the interferometer phase to cancel this contribution, which would correspond to $\textrm{mod}_{2\pi}(\phi_\mu)\sim 2.7$ radians. This was however not obvious from the phase scan in Fig. 2b in the main text, as in that measurement SiV was in fact initialized in the $\left|\downarrow\right\rangle_\textrm{\!\tiny{A}}\!$ state, inadvertently resulting in a different optical detuning of $\omega_\textrm{sbA}$ from the value in the correlation experiment. Thus, by the optimal phase from the interferometer phase scan in Fig. 2a in the main text, resulted in a sub-optimal phase setting for the actual correlation measurements. 
Consequently, choosing an optimal interferometer phase essentially has the same effect as an optimal sideband frequency, for the marginal error, namely to suppress the reflection from the $\left|\downarrow\downarrow\right\rangle_\textrm{\!\tiny{AB}}\!$ state. 

We further note that a further reduction of the carrier leakage is straight forward, e.g.\ using coherent canceling in a Sagnac configuration \cite{Li:14}, reducing the total marginal error budget of the heralded state errors to $\sim1\%$, related to the parameters of the cavity QED system. We are therefore optimistic that in future experiments, using devices with magentic defect concentrations more in line with our usual devices, a Bell state fidelity $>95\%$ can be reached on average. Using selected devices with superior coherence times, a Bell state fidelity $>97\%$ is expected to be possible.